\documentclass[11pt,a4paper]{article}
\usepackage[english]{babel}
\usepackage{amsmath,amsthm,amssymb,epsfig,latexsym}
\usepackage{color}
\usepackage{ulem}
\usepackage[numbers,square,sort&compress]{natbib}

\newcommand{\so}{ \alpha}
\newcommand{\st}{ \beta}
\newcommand{\sth}{ \gamma}
\newcommand{\qo}{\rho}
\newcommand{\qt}{\sigma}
\newcommand{\qth}{\tau}
\newcommand{\sz}{\mu}
\newcommand{\bsz}{\bar\mu}
\newcommand{\bu}{\bar u}
\newcommand{\bv}{\bar v}

\newcommand{\bw}{\bar w}

\newcommand{\bg}{\bar\gamma}
\newcommand{\prt}[1]{[#1]}
\newcommand{\be}[1]{\begin{equation}\label{#1}}
\newcommand{\ee}{\end{equation}}

\newtheorem{prop}{Proposition}[section]
\newtheorem{lemma}{Lemma}[section]
\newtheorem{cor}{Corollary}[section]
\newtheorem{Def}{Definition}[section]

 \makeatletter
 \@addtoreset{equation}{section}
 \makeatother

\newcommand{\bea}{\begin{eqnarray}}
\newcommand{\eea}{\end{eqnarray}}

\newcommand{\bT}{\mathbb{T}}

\begin{document}

\begin{center}
\begin{LARGE}
{\bf Multiple commutation relations\\in the models with $\mathfrak{gl}(2|1)$ symmetry}
\end{LARGE}

\vspace{40pt}

\begin{large}
{N.~A.~Slavnov\footnote{nslavnov@mi.ras.ru}}
\end{large}

 \vspace{12mm}

\vspace{4mm}

{\it Steklov Mathematical Institute,
Moscow, Russia}

\end{center}


\vspace{4mm}


\begin{abstract}
We consider quantum integrable models with $\mathfrak{gl}(2|1)$ symmetry. We derive
a set of multiple commutation relations between the monodromy matrix entries. These
multiple commutation relations allow us to obtain different representations for
Bethe vectors.
\end{abstract}

\vspace{1cm}

\centerline{{\bf Key words:} Bethe ansatz, monodromy matrix, commutation relations. }

\vspace{2mm}

\section{Introduction}

The algebraic Bethe ansatz is an efficient  method for finding the spectra of quantum Hamiltonians \cite{FadST79,FadT79,FadLH96}.
It is also known that this method is well suited for calculating correlation functions of quantum models
\cite{BogIK93L,KitMT00,KitKMST12,GohKS04}. The latest problem, within the framework of the algebraic Bethe ansatz reduces to the
calculating scalar products of Bethe vectors.
In their turn, the scalar products mentioned above appear to be a particular case of multiple commutation relations (MCR) of the
monodromy matrix entries. The subject of this paper are MCR in the $\mathfrak{gl}(2|1)$-based models.

In the models  solvable by the algebraic Bethe ansatz, the determining  is an $RRT$-relation
\begin{equation}\label{L-op-com}
R(u,v)\cdot (T(u)\otimes \mathbb{I})\cdot (\mathbb{I}\otimes T(v))=
(\mathbb{I}\otimes T(v))\cdot (T(u)\otimes \mathbb{I})\cdot R(u,v).
\end{equation}
Here $R(u,v)$ is an $R$-matrix acting in the tensor product $V\otimes V$ of an auxiliary vector spaces $V$. The monodromy matrix $T(u)$ acts
in $V\otimes \mathcal{H}$, where $\mathcal{H}$ is a Hilbert space of a quantum model. Equation \eqref{L-op-com} holds in the tensor
product $V\otimes V\otimes\mathcal{H}$.
It sets the commutation relations between the matrix elements $T_{ij}(u)$ of the monodromy matrix $T(u)$. At the same time, to calculate
scalar products and form factors one should know the MCR  of the type
\be{Intr-examp}
[T_{ij}(u_1)\dots T_{ij}(u_n),T_{kl}(v_1)\dots T_{kl}(v_m)].
\ee
Formally, knowing $[T_{ij}(u),T_{kl}(v)]$ one can calculate the commutator in \eqref{Intr-examp}. However, a straightforward use
of the $RTT$-relation in this case, leads to extremely complex expressions that hardly can be used for further analysis.

The MCR in the $\mathfrak{gl}(2)$-invariant models and their $q$-deformations are well known \cite{IzeK84,KitMT00,KitMST02}. Some generalizations
to the models with $\mathfrak{gl}(3)$-invariant $R$-matrix were considered in \cite{BelPRS12c}. In the present paper we deal with the models
based on  the superalgebra $\mathfrak{gl}(2|1)$.  Among the models described by this algebra, the most famous is the supersymmetric t-J model, which plays an important role in the condensed matter physics \cite{ZhaR88,Sch87}. The algebraic structure of this model as well as the application of the
algebraic Bethe ansatz were considered in \cite{BarB90,Sar91,BarBO91,Sar90,EssK92,FoeK93}.

As explained above, the MCR are needed for calculating the scalar products of Bethe vectors. However, in models with high
rank symmetry, to construct the Bethe vectors is already a non-trivial task. This problem was considered by different approaches
\cite{KhoP05,KhoPT,FraKPR09,BelPR10,TarV13,BelPRS12c}. A generalization of a trace formula for the Bethe vectors \cite{TarV13} for
the case of superalgebras $\mathfrak{gl}(m|n)$ was obtained in \cite{BelR08}. Recently, explicit expressions for the Bethe vectors
in the models with $\mathfrak{gl}(2|1)$ and $\mathfrak{gl}(1|2)$ symmetries were found in \cite{PakRS16a}. In this paper we use a method
of MCR and prove the equivalence of different representations for the $\mathfrak{gl}(2|1)$ Bethe vectors. We also obtain new
formulas for them.

The article is organized as follows. In section~\ref{S-BNN} we introduce the model under consideration and describe
necessary notation. Section~\ref{S-MCR-O} is devoted to the most simple MCR. In section~\ref{S-MCR-T} we formulate more complex
MCR, which allow us to prove different representations for the Bethe vectors in section~\ref{S-DRBV}. In sections~\ref{S-MCRa1},~\ref{S-MCRab} we prove the
MCR formulated in section~\ref{S-MCR-T}. In appendices~\ref{A-ID} and~\ref{A-SM} we gather some
identities necessary for the proof.

\section{Basic notions and notation\label{S-BNN}}

For $\mathfrak{gl}(2|1)$-based models an auxiliary vector space $V$ is a ${\mathbb Z}_2$-graded space ${\mathbb C}^{2|1}$  with a basis $\{{\rm e}_1,
{\rm e}_2,{\rm e}_3\}$. We call the vectors  $\{{\rm e}_1,{\rm e}_2\}$  even, while ${\rm e}_{3}$ is odd. Respectively, we  introduce  a party function on the set of indices as  $[1]=[2]=0$  and $\prt{3}=1$.

The $R$-matrix in \eqref{L-op-com} has the form
\begin{equation}\label{DYglmn}
R(u,v)\ =\  \mathbb{I}\otimes\mathbb{I}+g(u,v)\mathbb{P} \qquad g(u,v)=\frac{c}{u-v},
\end{equation}
where $c$ is a constant and $\mathbb{P}$ is a graded permutation matrix \cite{KulS80}. The tensor product in  \eqref{L-op-com}
also is graded leading to the set of commutation relations between the monodromy matrix entries $T_{ij}$:
\begin{equation}\label{TM-1}
[T_{ij}(u),T_{kl}(v)\}
=(-1)^{\prt{i}(\prt{k}+\prt{l})+\prt{k}\prt{l}}g(u,v)\Big(T_{kj}(v)T_{il}(u)-T_{kj}(u)T_{il}(v)\Big),
\end{equation}
where we have introduced a graded commutator as
\be{Def-SupC}
[T_{ij}(u),T_{kl}(v)\}= T_{ij}(u)T_{kl}(v) -(-1)^{(\prt{i}+\prt{j})(\prt{k}+\prt{l})}   T_{kl}(v)  T_{ij}(u).
\ee
Relabeling in \eqref{TM-1}  the subscripts $i\leftrightarrow k$,  $j\leftrightarrow l$, and replacing   $u\leftrightarrow v$ we  obtain one
more commutation relation
\begin{equation}\label{TM-2}
[T_{ij}(u),T_{kl}(v)\}=(-1)^{\prt{l}(\prt{i}+\prt{j})+\prt{i}\prt{j}}g(u,v)\Big(T_{il}(u)T_{kj}(v)-T_{il}(v)T_{kj}(u)\Big).
\end{equation}
Relations \eqref{TM-1}, \eqref{TM-2} are the starting point of our study.

Let us describe now the notation used below. We mostly use the same notation as we did in our papers concerning the
models with $\mathfrak{gl}(3)$ symmetry (see e.g. \cite{BelPRS12c}).
Apart from the function $g(x,y)$ we also introduce two functions
\be{univ-not}
 f(x,y)=1+g(x,y)=\frac{x-y+c}{x-y}, \qquad h(x,y)=\frac{f(x,y)}{g(x,y)}=\frac{x-y+c}{c}.
\ee
The functions introduced above have the following obvious properties:
 \be{propert}
 g(x,y)=-g(y,x),\qquad h(x,y)=\frac1{g(x,y-c)},\qquad  f(x-c,y)=\frac1{f(y,x)}.
 \ee

Extending the $\mathbb{Z}_2$-grading to the operators $T_{ij}$ by
\be{gradT}
[T_{ij}(u)]=[i]+[j],\quad \mod 2,
\ee
we distinguish between even operators (i.e. $[T_{ij}(u)]=0$) and odd  operators (i.e. $[T_{ij}(u)]=1$).
It follows from the commutation relations \eqref{TM-1}, \eqref{TM-2} that any even  operator $T_{ij}(u)$
commutes with itself for
arbitrary values of the argument $u$. On the other hand, one has for the odd operators $T_{ij}(u)$
\be{odd-comm}
\begin{aligned}
& h(v_1,v_2)T_{j3}(v_1)T_{j3}(v_2)=h(v_2,v_1)T_{j3}(v_1)T_{j3}(v_2),\\
&h(v_2,v_1)T_{3j}(v_1)T_{3j}(v_2)=h(v_1,v_2)T_{3j}(v_2)T_{3j}(v_1),
\end{aligned}
\qquad\qquad j=1,2.
\ee
Therefore it is convenient to introduce symmetric products of the odd operators.

\begin{Def}
Let $\bv=\{v_1,\dots,v_n\}$. Define
\be{bTc-def}
\bT_{j3}(\bv)= \frac{T_{j3}(v_1)\dots T_{j3}(v_n)}{\prod_{n\ge \ell>m\ge 1} h(v_\ell,v_m)}, \qquad j=1,2,
\ee
and
\be{bTr-def}
\bT_{3k}(\bv)= \frac{T_{3k}(v_1)\dots T_{3k}(v_n)}{\prod_{n\ge \ell>m\ge 1} h(v_m,v_\ell)}, \qquad k=1,2.
\ee
\end{Def}

Obviously, the operator products introduced above are symmetric over the parameters $\bv=\{v_1,\dots,v_n\}$.

Let us formulate a convention on the notation.
We  denote sets of variables by bar: $\bw$, $\bu$, $\bv$ etc. For the sake of generality, we will sometimes use such the notation, even in cases where
the set consists of only one element. Individual elements
of the sets are denoted by Latin subscripts: $w_j$, $u_k$ etc. The subsets complementary to the individual elements
are denoted by $\bu_k$: $\bu_k=\bu\setminus u_k$.  As a rule, the number of elements in the
sets is not shown explicitly in the equations, however we give these cardinalities in
special comments to the formulas. The notation $\bu\pm c$ means that all the elements of the set $\bu$ are
shifted by $\pm c$:  $\bu\pm c=\{u_1\pm c,\dots,u_n\pm c\}$.

 Subsets of variables (including subsets consisting of one element) are labeled by Greek subscripts: $\bu_{\so}$, $\bv_{\qt}$, $\bw_{\st}$ etc.
 The union of the sets is denoted by braces: $\bw=\{\bu,\bv\}$.
A subset complementary to the set $\bu_{\so}$ is denoted by $\bu_{\bar\so}$: $\bu_{\bar\so}=\bu\setminus\bu_{\so}$.
The notation $\bu\Rightarrow\{\bu_{\so},\bu_{\bar\so}\}$ means that the
set $\bu$ is divided into two  subsets $\bu_{\so}$ and $\bu_{\bar\so}$. Similarly, the notation
$\bv\Rightarrow\{\bv_{\so},\bv_{\st},\bv_{\sth}\}$ means that the
set $\bv$ is divided into three disjoint  subsets. Hereby, $\{\bv_{\so},\bv_{\st}\}=\bv_{\bar\sth}$,
$\{\bv_{\so},\bv_{\sth}\}=\bv_{\bar\st}$, and $\{\bv_{\sth},\bv_{\st}\}=\bv_{\bar\so}$.
We assume that the elements in every subset of variables are ordered in such a way that the sequence of
their subscripts is strictly increasing. We call this ordering  natural order.


In order to avoid too cumbersome formulas we use shorthand notation for products of commuting operators.
Similarly to \eqref{bTc-def}, \eqref{bTr-def} we introduce
 \be{SH-prod0}
 T_{ij}(\bu)=\prod_{u_k\in\bu}  T_{ij}(u_k),\qquad \text{for}\quad \prt{i}+\prt{j}=0,\quad \mod 2.
 \ee
One should follow the same prescription for the products of the functions $g$, $f$, $h$. Namely, if such the function depends
on a set of variables, this means that one should take the product over the corresponding set.
For example,
 \be{SH-prod}
 h(v, \bu)= \prod_{u_j\in\bu} h(v, u_j);\quad
 f(\bu,\bv)=\prod_{u_j\in\bu}\prod_{v_k\in\bv} f(u_j,v_k);\quad
  g(\bv_{\so},\bv_{\bar\so})=\prod_{v_j\in\bv_{\so}}\prod_{v_k\in\bv_{\bar\so}} g(v_j,v_k).
 \ee

To conclude this section we introduce a partition function of the six-vertex model with domain wall boundary conditions (DWPF)  $K(\bu|\bv)$ \cite{Kor82,Ize87}.
This function plays an important role in
the MCR.  It depends on two sets of variables $\bu$ and $\bv$, such that
$\#\bu=\#\bv$. By definition $K(\emptyset|\emptyset)=1$. Otherwise, if $\#\bu=\#\bv=n$, $n>0$, then the function $K(\bu|\bv)$ has the following determinant representation:
\begin{equation}\label{K-def}
K(\bu|\bv)
=\Delta'(\bu)\Delta(\bv)h(\bu,\bv)
\det_n \left(\frac{g(u_j,v_k)}{h(u_j,v_k)}\right),
\end{equation}
where $\Delta'(\bu)$ and $\Delta(\bv)$ are
\be{def-Del}
\Delta'(\bu)=
\prod_{1\le j<k\le n} g(u_j,u_k),\qquad {\Delta}(\bv)= \prod_{n\ge j>k\ge 1}g(v_j,v_k).
\ee
It is easy to see that $K(\bu|\bv)$ is symmetric over $\bu$ and symmetric over $\bv$, however  $K(\bu|\bv)\ne
 K(\bv|\bu)$. Some useful properties of the DWPF are gathered in appendix~\ref{A-ID}.

\section{MCR of operators belonging to the same row or column \label{S-MCR-O}}

In this section we consider relatively simple MCR. They occur if the
operators $T_{ij}$ and $T_{kl}$ belong to the same row or column of the monodromy matrix, i.e.
either $i=k$ or $j=l$.

\begin{prop}\label{P-ijik}
Let $\#\bu=n$,  $\#\bv=m$, and $\{\bu,\bv\}=\bw$. Then for $i,j,k<3$
\be{Mult-comTijTik}
T_{ij}(\bu)T_{ik}(\bv)=(-1)^{n}\sum K(\bw_{\so}|\bu+c)f(\bw_{\bar\so},\bw_{\so})
T_{ik}(\bw_{\bar\so})T_{ij}(\bw_{\so}).
\ee
For $i,j<3$
\be{Mult-comTi3Tj3}
\bT_{i3}(\bu)\bT_{j3}(\bv)=(-1)^n h(\bv,\bu)\sum K(\bu|\bw_{\so}+c)g(\bw_{\so},\bw_{\bar\so})\bT_{j3}(\bw_{\bar\so})\bT_{i3}(\bw_{\so}),
\ee
\be{TijTi3-0}
T_{ij}(\bu)\bT_{i3}(\bv)=\sum h(\bw_{\bar\so},\bu)g(\bw_{\bar\so},\bw_{\so})\bT_{i3}(\bw_{\bar\so})T_{ij}(\bw_{\so}),
\ee
\be{Mult-comTi3Tij}
\bT_{i3}(\bu)T_{ij}(\bv)=\sum h(\bv,\bw_{\so})g(\bw_{\bar\so},\bw_{\so})T_{ij}(\bw_{\bar\so})\bT_{i3}(\bw_{\so}).
\ee
For $i<3$
\be{T33Ti3-0}
T_{33}(\bu)\bT_{3i}(\bv)=\sum h(\bu,\bw_{\bar\so}) g(\bw_{\so},\bw_{\bar\so})\bT_{3i}(\bw_{\bar\so})T_{33}(\bw_{\so}).
\ee
\be{Mult-comTi3T33}
\bT_{3i}(\bu)T_{33}(\bv)=\sum   h(\bw_{\so},\bv)g(\bw_{\so},\bw_{\bar\so})T_{33}(\bw_{\bar\so})\bT_{3i}(\bw_{\so}),
\ee
All the sums are taken over partitions $\bw\Rightarrow\{\bw_{\so},\bw_{\bar\so}\}$  with
$\#\bw_{\so}=n$ and $\#\bw_{\bar\so}=m$.
\end{prop}

Proposition~\ref{P-ijik} gives us the MCR for the operators belonging to the same row of the
monodromy matrix. In order to obtain the commutation relations for the operators from the same column it is enough to
apply antimorphism of the algebra \eqref{L-op-com}
\cite{PakRS16a}
\be{psi}
\psi\bigl(T_{ij}(u)\bigr)=(-1)^{[i][j]+[i]}T_{ji}(u),\qquad
\psi\bigl(T_{ij}(u)T_{kl}(v)\bigr)=(-1)^{[T_{ij}][T_{kl}]}\psi\bigl(T_{kl}(v)\bigr)\psi\bigl(T_{ji}(u)\bigr).
\ee
Clearly, that after the action of $\psi$ onto  equations \eqref{Mult-comTijTik}--\eqref{Mult-comTi3T33} we obtain
MCR for the operators belonging to the same column of the monodromy matrix.

The proofs of \eqref{Mult-comTijTik}--\eqref{T33Ti3-0} are similar one to each other. They use induction over the
numbers of the operators and summation lemmas~\ref{main-ident-C}, \ref{main-ident}. The proof of \eqref{Mult-comTijTik}
(based on lemma~\ref{main-ident}) was given in \cite{BelPRS12c}. In order to show how lemma~\ref{main-ident-C} works we
consider the proof of \eqref{TijTi3-0}.

It follows  from the commutation relations \eqref{TM-2} that
\be{TijTi3one}
T_{ij}(u)T_{i3}(v)=f(v,u)T_{i3}(v)T_{ij}(u)+g(u,v)T_{i3}(u)T_{ij}(v).
\ee
It is easy to see that this equation coincides with \eqref{TijTi3-0} for $n=m=1$.  Then one can use double induction over
$n$ and $m$. However, for $n=1$ or $m=1$ it is better to apply the standard scheme of the algebraic Bethe ansatz. Let for
definiteness $n=1$ and $m>1$. Then moving the operator  $T_{ij}(u)$ through the product $\bT_{i3}(\bv)$ we obtain the
following result
\be{TijTi3many}
T_{ij}(u)\bT_{i3}(\bv)=\Lambda \bT_{i3}(\bv)T_{ij}(u)+\sum_{k=1}^m\Lambda_k \bT_{i3}(\{\bv_k,u\})T_{ij}(v_k).
\ee
Here $\Lambda$ and $\Lambda_k$ are some rational coefficients to be determined, and we recall that $\bv_k=\bv\setminus v_k$. Obviously, in order to
obtain the term proportional to $\Lambda$, the operator $T_{ij}(u)$ should go to the right preserving its
original argument. This immediately gives
\be{La}
\Lambda=f(\bv,u).
\ee

Let us find now $\Lambda_1$. The  product $\bT_{i3}(\{\bv_1,u\})T_{ij}(v_1)$ arises if and only if
the operators $T_{ij}(u)$ and $T_{i3}(v_1)$ exchange their arguments:
\be{Lak-1}
T_{ij}(u)\bT_{i3}(\bv)=T_{ij}(u)T_{i3}(v_1)\frac{\bT_{i3}(\bv_1)}{h(\bv_1,v_1)}=
g(u,v_1)T_{i3}(u)T_{ij}(v_1)\frac{\bT_{i3}(\bv_1)}{h(\bv_1,v_1)}+UWT,
\ee
where $UWT$ means {\it unwanted terms}. Generically, we call a term unwanted if it does not contribute to the desirable coefficient.
In the case under consideration, the term proportional to $T_{i3}(v_1)T_{ij}(u)$ is unwanted, because moving the operator $T_{ij}(u)$
through the product $\bT_{i3}(\bv_1)$ we cannot obtain $T_{ij}(v_1)$ in the extreme right position.

Now, when the operator $T_{ij}(v_1)$ moves  to the right it should preserve its argument $v_1$. This gives us
\be{Lak-2}
T_{ij}(u)\bT_{i3}(\bv)=
g(u,v_1)g(\bv_1,v_1)T_{i3}(u)\bT_{i3}(\bv_1)T_{ij}(v_1)+UWT,
\ee
where we used $g(\bv_1,v_1)=f(\bv_1,v_1)/h(\bv_1,v_1)$.
It remains to transform the product $T_{i3}(u)\bT_{i3}(\bv_1)$ into $\bT_{i3}(\{\bv_1,u\})$, and we finally obtain
\be{Lak-3}
\Lambda_1=g(u,v_1)g(\bv_1,v_1)h(\bv_1,u).
\ee
Due to the symmetry of $\bT_{i3}(\bv)$ over $\bv$ we conclude that
\be{Lak-4}
\Lambda_k=g(u,v_k)g(\bv_k,v_k)h(\bv_k,u).
\ee
Thus,
\be{TijTi3many-1}
T_{ij}(u)\bT_{i3}(\bv)=f(\bv,u) \bT_{i3}(\bv)T_{ij}(u)+\sum_{k=1}^mg(u,v_k)g(\bv_k,v_k)h(\bv_k,u) \bT_{i3}(\{\bv_k,u\})T_{ij}(v_k).
\ee
Comparing equations \eqref{TijTi3many-1} and \eqref{T33Ti3-0} at $n=1$ we see that the first term in \eqref{TijTi3many-1} corresponds
to the partition $\bw_{\so}=u$ and $\bw_{\bar\so}=\bv$. The other terms in \eqref{TijTi3many-1} appear if we set $\bw_{\so}=v_k$ and
$\bw_{\bar\so}=\{\bv_k,u\}$. Thus, \eqref{T33Ti3-0} is proved for $n=1$ and $m$ arbitrary.

Generalization for $n>1$ can be done via induction over $n$. Suppose that \eqref{T33Ti3-0} holds for some $n-1$. Then acting
successively with $T_{ij}(\bu_n)$ and $T_{ij}(u_n)$ on the product $\bT_{i3}(\bv)$ we obtain
\be{TijTi3-1}
T_{ij}(u_n)T_{ij}(\bu_n)\bT_{i3}(\bv)=\sum h(\bw_{\bar\so},\bu_n)g(\bw_{\bar\so},\bw_{\so})
h(\bw_{\qt},u_n)g(\bw_{\qt},\bw_{\qo})\bT_{i3}(\bw_{\qt})T_{ij}(\bw_{\qo})T_{ij}(\bw_{\so}).
\ee
Here we first have partitions of the set
$\{\bu_n,\bv\}$ into subsets $\bw_{\so}$ and $\bw_{\bar\so}$ with $\#\bw_{\so}=n-1$ and $\#\bw_{\bar\so}=m$. Then we combine $u_n$ and $\bw_{\bar\so}$ and divide this
set into subsets $\bw_{\qo}$ and $\bw_{\qt}$ with $\#\bw_{\qo}=1$ and $\#\bw_{\qt}=m$. As a result we have divided the set
$\bw=\{\bu,\bv\}$ into three subsets $\{\bw_{\so},\bw_{\qo},\bw_{\qt}\}$ in such a way that  $u_n\notin\bw_{\so}$. Substituting
$\bw_{\bar\so}=\{\bw_{\qo},\bw_{\qt}\}\setminus u_n$ into \eqref{TijTi3-1} we arrive at
\begin{multline}\label{TijTi3-2}
T_{ij}(\bu)\bT_{i3}(\bv)=\sum \frac{h(\bw_{\qo},\bu_n)h(\bw_{\qt},\bu_n)g(\bw_{\qo},\bw_{\so})g(\bw_{\qt},\bw_{\so})}
{h(u_n,\bu_n)g(u_n,\bw_{\so}) }\\
\times h(\bw_{\qt},u_n)g(\bw_{\qt},\bw_{\qo})\bT_{i3}(\bw_{\qt})T_{ij}(\bw_{\qo})T_{ij}(\bw_{\so}).
\end{multline}
Here the sum is taken over partitions of the set
$\{\bu,\bv\}$ into  subsets $\{\bw_{\so},\bw_{\qo},\bw_{\qt}\}$. Observe that we have got rid of the restriction $u_n\notin\bw_{\so}$.
Indeed, due to the factor $g(u_n,\bw_{\so})$ in the denominator all the terms of the sum vanish as soon as $u_n\in\bw_{\so}$.

Let us  set $\{\bw_{\so},\bw_{\qo}\}=\bw_{\sz}$. Then $\bw_{\qt}=\bw_{\bar\sz}$, and \eqref{TijTi3-2} turns into
\begin{equation}\label{TijTi3-3}
T_{ij}(\bu)\bT_{i3}(\bv)=\sum \frac{h(\bw_{\bar\sz},\bu)h(\bw_{\sz},\bu_n)g(\bw_{\bar\sz},\bw_{\sz})}
{h(u_n,\bu_n)g(u_n,\bw_{\sz}) } \bT_{i3}(\bw_{\bar\sz})T_{ij}(\bw_{\sz})
 \left(g(\bw_{\qo},\bw_{\so})\frac{g(u_n,\bw_{\qo})}{h(\bw_{\so},\bu_n)}\right).
\end{equation}
One can say that here we first divide the set $\{\bu,\bv\}$ into  subsets $\{\bw_{\sz},\bw_{\bar\sz}\}$, and then
divide the subset $\bw_{\sz}$ into $\bw_{\so}$ and $\bw_{\qo} $. The sum over partitions
$\bw_{\sz}\Rightarrow\{\bw_{\so},\bw_{\qo}\} $ (see the terms in the parenthesis in \eqref{TijTi3-3}) can be computed explicitly via
lemma~\ref{main-ident-C}
\be{sum-w0}
\sum g(\bw_{\qo},\bw_{\so})\frac{g(u_n,\bw_{\qo})}{h(\bw_{\so},\bu_n)}=
-\sum g(\bw_{\qo},\bw_{\so})g(u_n,\bw_{\qo})g(\bw_{\so},\bu_n-c)=
\frac{h(u_n,\bu_n)g(u_n,\bw_{\sz}) } {h(\bw_{\sz},\bu_n)}.
\ee
Substituting this into \eqref{TijTi3-3} we finally arrive at
\be{TijTi3-00}
T_{ij}(\bu)\bT_{i3}(\bv)=\sum h(\bw_{\bar\sz},\bu)g(\bw_{\bar\sz},\bw_{\sz})\bT_{i3}(\bw_{\bar\sz})T_{ij}(\bw_{\sz}),
\ee
what coincides with \eqref{TijTi3-0} at $\#\bu=n$ up to the labels of the subsets.

Other equations of proposition~\ref{P-ijik} can be proved in the similar manner.

\section{MCR of operators belonging to different rows and columns \label{S-MCR-T}}

MCR of the operators belonging to the different rows and columns of the monodromy matrix are
much more sophisticated than the ones considered above. We will consider only one specific example of these MCR.
This example is important, because it allows one to obtain different representations for the
Bethe vectors.

\begin{Def}
For $\#\bu=a$ and $\#\bv=b$ define two operators
\be{TPhi-expl1}
X_{a,b}(\bu,\bv)=\sum g(\bv_{\so},\bu_{\so}) f(\bu_{\so},\bu_{\bar\so}) g(\bv_{\bar\so},\bv_{\so})h(\bu_{\so},\bu_{\so})\;
\bT_{13}(\bu_{\so})\,T_{12}(\bu_{\bar\so})\,\bT_{23}(\bv_{\bar\so})\,T_{22}(\bv_{\so}),
\ee
and
\be{TPhi-expl2}
Y_{a,b}(\bu,\bv)=\sum K(\bv_{\so}|\bu_{\so}) f(\bu_{\so},\bu_{\bar\so}) g(\bv_{\bar\so},\bv_{\so})\;
\bT_{13}(\bv_{\so})\,\bT_{23}(\bv_{\bar\so})\,T_{12}(\bu_{\bar\so})\,T_{22}(\bu_{\so}).
\ee
In both equations the sum is taken over partitions $\bv\Rightarrow\{\bv_{\so},\bv_{\bar\so}\}$ and $\bu\Rightarrow\{\bu_{\so},\bu_{\bar\so}\}$ with the restriction $\#\bu_{\so}=\#\bv_{\so}=n$, where $n=0,1,\dots,\min(a,b)$.
\end{Def}

The operator \eqref{TPhi-expl1} was considered in \cite{PakRS16a}. There it was proved that it possesses the following recursion:
\begin{multline}\label{recurs-Phi}
X_{a,b}(\bu,\bv) = T_{12}(u_a)\, X_{a-1,b}(\bu_a,\bv)\\
+\sum g(\bv_{\qo},u_a)f(\bv_{\qo},\bu_a)g(\bv_{\bar\qo},\bv_{\qo})
\,T_{13}(u_a) X_{a-1,b-1}(\bu_a,\bv_{\bar\qo})T_{22}(\bv_{\qo}).
\end{multline}
Here the sum is taken over partitions $\bv=\{\bv_{\qo},\bv_{\bar\qo}\}$, where the subset $\bv_{\qo}$ consists of one element.
Recall that $\bu_a=\bu\setminus u_a$.

\begin{prop}\label{P-MP}
The operator \eqref{TPhi-expl2} satisfies recursion \eqref{recurs-Phi}
\begin{multline}\label{recurs-Y}
Y_{a,b}(\bu,\bv) = T_{12}(u_a)\, Y_{a-1,b}(\bu_a,\bv) \\
+\sum g(\bv_{\qo},u_a)f(\bv_{\qo},\bu_a)g(\bv_{\bar\qo},\bv_{\qo})
\,T_{13}(u_a) Y_{a-1,b-1}(\bu_a,\bv_{\bar\qo})T_{22}(\bv_{\qo}).
\end{multline}
The notation is the same as in \eqref{recurs-Phi}.
\end{prop}

The proof of this proposition will be given in the following sections. Here we would like to
mention only that proposition~\ref{P-MP} yields
\be{Y-Xab}
Y_{a,b}(\bu,\bv)=X_{a,b}(\bu,\bv),\qquad \forall~ a,\,b.
\ee
Indeed, it is clear that
\be{Y-Phi0}
Y_{0,b}(\emptyset,\bv)=X_{0,b}(\emptyset,\bv)=\bT_{23}(\bv),\qquad \forall~ b.
\ee
Then using recursions \eqref{recurs-Phi} and \eqref{recurs-Y} we arrive at \eqref{Y-Xab}.

Equation \eqref{Y-Xab} can be considered as MCR of the operators
$T_{12}(\bu)$ and $\bT_{23}(\bv)$. Indeed, we have
\be{XY-TT}
\begin{aligned}
X_{a,b}(\bu,\bv)&=T_{12}(\bu)\,\bT_{23}(\bv)+\dots,\\
Y_{a,b}(\bu,\bv)&=\bT_{23}(\bv)\,T_{12}(\bu)+\dots,
\end{aligned}
\ee
where dots mean the terms proportional to the operators $T_{13}$ and $T_{22}$. Thus, equation \eqref{Y-Xab} provides us with
a multiple commutator $[T_{12}(\bu),\,\bT_{23}(\bv)]$ in terms of the operators $T_{13}$, $T_{22}$, and the products of
$T_{12}$ and $T_{23}$ with less number of arguments.

\section{Different representations for Bethe vectors\label{S-DRBV}}

MCR \eqref{Y-Xab} allow us to find different representations for the
Bethe vectors in the models with $\mathfrak{gl}(2|1)$ symmetry. Recall that Bethe vectors are special polynomials in
operators $T_{ij}(u)$ with $i\le j$ applied to the pseudovacuum vector $\Omega$. This vector possesses the following
properties:
\be{vac}
\begin{aligned}
T_{ii}(u)\Omega&=\lambda_i(u)\Omega,\\
T_{ij}(u)\Omega&=0,\qquad i>j.
\end{aligned}
\ee
Here $\lambda_i(u)$ are some scalar functions depending on a specific model.

The Bethe vectors  depend on two sets of variables (Bethe parameters) $\bu$ and $\bv$. We denote the Bethe vectors $\Phi_{a,b}(\bu;\bv)$.
It was proved in \cite{PakRS16a} that they have the following explicit form:
$\Phi_{a,b}(\bu;\bv)=X_{a,b}(\bu;\bv)\Omega$, where $X_{a,b}$ is given by \eqref{TPhi-expl1}. As we have proved that
$X_{a,b}(\bu;\bv)=Y_{a,b}(\bu;\bv)$, we immediately obtain an alternative representation $\Phi_{a,b}(\bu;\bv)=Y_{a,b}(\bu;\bv)\Omega$,
where $Y_{a,b}$ is given by \eqref{TPhi-expl2}. Thus, we have the following explicit formulas for the Bethe vectors:
\be{BV-1}
\begin{aligned}
\Phi_{a,b}(\bu,\bv)&=\sum g(\bv_{\so},\bu_{\so}) f(\bu_{\so},\bu_{\bar\so}) g(\bv_{\bar\so},\bv_{\so})h(\bu_{\so},\bu_{\so})\;
\bT_{13}(\bu_{\so})\,T_{12}(\bu_{\bar\so})\,\bT_{23}(\bv_{\bar\so})\,\lambda_{2}(\bv_{\so})\Omega,\\
\Phi_{a,b}(\bu,\bv)&=\sum K(\bv_{\so}|\bu_{\so}) f(\bu_{\so},\bu_{\bar\so}) g(\bv_{\bar\so},\bv_{\so})\;
\bT_{13}(\bv_{\so})\,\bT_{23}(\bv_{\bar\so})\,T_{12}(\bu_{\bar\so})\,\lambda_{2}(\bu_{\so})\Omega.
\end{aligned}
\ee
Here we have used that $T_{22}(u)\Omega=\lambda_{2}(u)\Omega$ and extended the convention on the shorthand notation \eqref{SH-prod}
to the products of the functions $\lambda_{2}(u)$.

Due to proposition~\ref{P-ijik} we can easily obtain another representations
\be{BV-2}
\begin{aligned}
\Phi_{a,b}(\bu,\bv)&=\sum g(\bv_{\so},\bu_{\so}) f(\bu_{\bar\so},\bu_{\so}) g(\bv_{\bar\so},\bv_{\so})
f(\bv_{\so},\bu_{\bar\so})h(\bu_{\so},\bu_{\so})\\
&\hspace{60mm}\times
T_{12}(\bu_{\bar\so})\,\bT_{13}(\bu_{\so})\,\bT_{23}(\bv_{\bar\so})\,\lambda_{2}(\bv_{\so})\Omega,\\
\Phi_{a,b}(\bu,\bv)&=\sum K(\bv_{\so}|\bu_{\so}) f(\bu_{\so},\bu_{\bar\so})g(\bv_{\bar\so},\bv_{\so})f(\bv_{\bar\so},\bu_{\so})\;
\bT_{23}(\bv_{\bar\so})\,\bT_{13}(\bv_{\so})\,T_{12}(\bu_{\bar\so})\,\lambda_{2}(\bu_{\so})\Omega.
\end{aligned}
\ee
Let us prove, for instance, the second equation \eqref{BV-2}. Consider the sum over partitions
$\bv\Rightarrow \{\bv_{\so},\bv_{\bar\so}\}$ in the second equation \eqref{BV-1} for $n=\#\bv_{\so}$ fixed. Using \eqref{Mult-comTi3Tj3}
we obtain
\begin{multline}\label{A-rep1}
\sum K(\bv_{\so}|\bu_{\so}) g(\bv_{\bar\so},\bv_{\so})\;
\bT_{13}(\bv_{\so})\,\bT_{23}(\bv_{\bar\so})\\
=(-1)^n\sum K(\bv_{\so}|\bu_{\so}) f(\bv_{\bar\so},\bv_{\so})K(\bv_{\so}|\bv_{\st}+c)\;g(\bv_{\bar\st},\bv_{\st})
\bT_{23}(\bv_{\bar\st})\,\bT_{13}(\bv_{\st}),
\end{multline}
where we have an additional sum over partitions $\bv\Rightarrow \{\bv_{\st},\bv_{\bar\st}\}$.
The partitions $\bv\Rightarrow \{\bv_{\so},\bv_{\bar\so}\}$ and  $\bv\Rightarrow \{\bv_{\st},\bv_{\bar\st}\}$ are independent
except that $\#\bv_{\so}=\#\bv_{\st}=n$. Thus, we can take the sum over partitions $\bv\Rightarrow \{\bv_{\so},\bv_{\bar\so}\}$
in the r.h.s. of \eqref{A-rep1}. For this we first transform the DWPF $K(\bv_{\so}|\bv_{\st}+c)$ using \eqref{K-K}
\begin{multline}\label{K-trans}
K(\bv_{\so}|\bv_{\st}+c)=(-1)^{b-n}K(\{\bv_{\so},\bv_{\bar\so}\}|\{\bv_{\st}+c,\bv_{\bar\so}+c\})\\
=(-1)^{b-n}K(\{\bv_{\st},\bv_{\bar\st}\}|\{\bv_{\st}+c,\bv_{\bar\so}+c\})=
(-1)^{b}K(\bv_{\bar\st}-c|\bv_{\bar\so}).
\end{multline}
Substituting this into \eqref{A-rep1} we obtain
\begin{multline}\label{A-rep2}
(-1)^n\sum K(\bv_{\so}|\bu_{\so}) f(\bv_{\bar\so},\bv_{\so})K(\bv_{\so}|\bv_{\st}+c)=
(-1)^{b-n}\sum K(\bv_{\so}|\bu_{\so}) f(\bv_{\bar\so},\bv_{\so})K(\bv_{\bar\st}-c|\bv_{\bar\so})\\
=(-1)^{b} f(\bv,\bu_{\so})K(\{\bu_{\so}-c,\bv_{\bar\st}-c\}|\bv)=f(\bv_{\bar\st},\bu_{\so})K(\bv_{\st}|\bu_{\so}),
\end{multline}
where we have used \eqref{Sym-Part-old1} and \eqref{K-K}. Substituting this into  \eqref{A-rep1} we arrive at
\begin{equation}\label{A-rep3}
\sum K(\bv_{\so}|\bu_{\so}) g(\bv_{\bar\so},\bv_{\so})\;
\bT_{13}(\bv_{\so})\,\bT_{23}(\bv_{\bar\so})
=\sum K(\bv_{\st}|\bu_{\so})\;f(\bv_{\bar\st},\bu_{\so})g(\bv_{\bar\st},\bv_{\st})
\bT_{23}(\bv_{\bar\st})\,\bT_{13}(\bv_{\st}).
\end{equation}
It remains to replace here $\bv_{\st}\to\bv_{\so}$ and $\bv_{\bar\st}\to\bv_{\bar\so}$, and we obtain the second representation
\eqref{BV-2}.

Similarly, starting with the first representation \eqref{BV-1} and using summation formula \eqref{Sym-Part-new1} we find the
first representation \eqref{BV-2}.

To conclude this section we give also explicit representations for the dual Bethe vectors. They belong to the dual space and can be
obtained from \eqref{BV-1}, \eqref{BV-2} via antimorphism \eqref{psi}. This antimorphism sends the pseudovacuum vector to its
dual: $\psi(\Omega)=\Omega^\dagger$. The vector $\Omega^\dagger$ belongs to the dual space and
possesses the properties
\be{dvac}
\begin{aligned}
\Omega^\dagger T_{ii}(u)&=\lambda_i(u)\Omega^\dagger,\\
\Omega^\dagger T_{ij}(u)&=0,\qquad i<j.
\end{aligned}
\ee
Here the functions $\lambda_i(u)$ are the same as in \eqref{vac}.

Acting with $\psi$ onto \eqref{BV-1}, \eqref{BV-2} we find
\be{dBV-1}
\begin{aligned}
\Phi^\dagger_{a,b}(\bu,\bv)&=(-1)^{b(b-1)/2}\sum g(\bv_{\so},\bu_{\so}) f(\bu_{\so},\bu_{\bar\so}) g(\bv_{\bar\so},\bv_{\so})h(\bu_{\so},\bu_{\so})\\
&\hspace{60mm}\times \lambda_{2}(\bv_{\so})\Omega^\dagger \,\bT_{32}(\bv_{\bar\so})\,T_{21}(\bu_{\bar\so})\bT_{31}(\bu_{\so}),\\
\Phi^\dagger_{a,b}(\bu,\bv)&=(-1)^{b(b-1)/2}\sum K(\bv_{\so}|\bu_{\so}) f(\bu_{\so},\bu_{\bar\so}) g(\bv_{\bar\so},\bv_{\so})\\
&\hspace{60mm}\times \lambda_{2}(\bu_{\so})\Omega^\dagger  \,T_{21}(\bu_{\bar\so})\,\bT_{32}(\bv_{\bar\so})\bT_{31}(\bv_{\so}),
\end{aligned}
\ee
and
\be{dBV-2}
\begin{aligned}
\Phi^\dagger_{a,b}(\bu,\bv)&=(-1)^{b(b-1)/2}\sum g(\bv_{\so},\bu_{\so}) f(\bu_{\bar\so},\bu_{\so}) g(\bv_{\bar\so},\bv_{\so})
f(\bv_{\so},\bu_{\bar\so})h(\bu_{\so},\bu_{\so})\\
&\hspace{60mm}\times \lambda_{2}(\bv_{\so})\Omega^\dagger \,\bT_{32}(\bv_{\bar\so})
\,\bT_{31}(\bu_{\so})T_{21}(\bu_{\bar\so}),\\
\Phi^\dagger_{a,b}(\bu,\bv)&=(-1)^{b(b-1)/2}\sum K(\bv_{\so}|\bu_{\so}) f(\bu_{\so},\bu_{\bar\so})g(\bv_{\bar\so},\bv_{\so})f(\bv_{\bar\so},\bu_{\so})\\
&\hspace{60mm}\times
\,\lambda_{2}(\bu_{\so})\Omega^\dagger  \,T_{21}(\bu_{\bar\so})\bT_{31}(\bv_{\so})\,\bT_{32}(\bv_{\bar\so}).
\end{aligned}
\ee

\section{MCR \eqref{Y-Xab} for $a=1$ \label{S-MCRa1}}

For $a=1$ equation \eqref{Y-Xab} takes the form
\begin{equation}\label{Mult-T12T230}
[T_{12}(u),\bT_{23}(\bv)]=\sum g(u,\bv_{\so}) g(\bv_{\bar\so},\bv_{\so})\Bigl(T_{13}(u)\bT_{23}(\bv_{\bar\so})\,T_{22}(\bv_{\so})-
T_{13}(\bv_{\so})\,\bT_{23}(\bv_{\bar\so})\,T_{22}(u)\Bigr).
\end{equation}
Here the sum is taken over partitions of the set $\bv$ into subsets $\bv_{\so}$ and $\bv_{\bar\so}$ with a restriction $\#\bv_{\so}=1$.
This MCR can be proved via the standard method of the algebraic Bethe ansatz. Indeed, due to  \eqref{TM-2} we have
\begin{multline}\label{Mult-T12T231}
[T_{12}(u),\bT_{23}(\bv)]=\prod_{1\le j<l\le b}\frac1{h(v_j,v_l)}\sum_{k=1}^b g(u,v_{k})\\
\times T_{23}(v_1)\dots T_{23}(v_{k-1})\bigl(T_{13}(u)T_{22}(v_k)-
T_{13}(v_k)T_{22}(u)\bigr)T_{23}(v_{k+1})\dots T_{23}(v_{b}).
\end{multline}
Then moving the operator $T_{13}$ to the left and the operator $T_{22}$ to the right we eventually obtain the following result
\begin{multline}\label{Mult-T12T232}
[T_{12}(u),\bT_{23}(\bv)]=\sum_{k=1}^b \Bigl(\Lambda_k T_{13}(u)\bT_{23}(\bv_{k})\,T_{22}(v_k)+\tilde \Lambda_k
T_{13}(v_k)\,\bT_{23}(\bv_{k})\,T_{22}(u)\Bigr)\\
+\sum_{\substack{j,k=1\\ j\ne k}}^b M_{jk} T_{13}(v_j)\bT_{23}(\{\bv_{j,k},u\})\,T_{22}(v_k).
\end{multline}
Here $\Lambda_k$, $\tilde \Lambda_k$, and $M_{jk}$ are some rational coefficients. The set $\bv_{j,k}$ is $\bv\setminus\{v_j,v_k\}$. Due
to the symmetry of $\bT_{23}(\bv)$ over $\bv$ it is enough to find $\Lambda_1$, $\tilde \Lambda_b$, and $M_{21}$.
For this we need particular cases of equations \eqref{Mult-comTijTik}, \eqref{Mult-comTi3Tj3}
\begin{equation}\label{Mult-T22T12}
T_{22}(v)T_{12}(\bu)=f(v,\bu)T_{12}(\bu)T_{22}(v)+\sum g(\bu_{\qo},v) f(\bu_{\qo},\bu_{\bar\qo})T_{12}(v)T_{12}(\bu_{\bar\qo})T_{22}(\bu_{\qo}),
\end{equation}
\begin{equation}\label{Mult-T23T13}
\bT_{23}(\bu)T_{13}(v)=(-1)^b f(v,\bu)T_{13}(v)\bT_{23}(\bu)+\sum g(v,\bu_{\qo}) g(\bu_{\bar\qo},\bu_{\qo})h(v,\bu_{\bar\qo})
T_{13}(\bv_{\qo})\bT_{23}(\{v,\bu_{\bar\qo}\}).
\end{equation}
Here the sums are taken over partitions $\bu\Rightarrow\{\bu_{\qo},\bu_{\bar\qo}\}$ with $\#\bu_{\qo}=1$.

Let us find $\Lambda_1$. Then
it is clear that only the term $k=1$ in \eqref{Mult-T12T231} contributes into this coefficient. Indeed, all other terms cannot produce $T_{22}(v_1)$ in the  extreme  right position.
Thus, presenting $\bT_{23}(\bv)$ as
\begin{equation}\label{Lam-10}
\bT_{23}(\bv)=T_{23}(v_1)\frac{\bT_{23}(\bv_{1})}{h(\bv_1,v_1)},
\end{equation}
we obtain
\begin{equation}\label{Lam-1}
[T_{12}(u),\bT_{23}(\bv)]=g(u,v_1)\bigl(T_{13}(u)T_{22}(v_1)-
T_{13}(v_1)T_{22}(u)\bigr)\frac{\bT_{23}(\bv_{1})}{h(\bv_1,v_1)}+UWT.
\end{equation}
Clearly, the product $T_{13}(v_1)T_{22}(u)$ does not contribute to the desirable coefficient. In the remaining product $T_{13}(u)T_{22}(v_1)$
the operator $T_{13}(u)$ should be kept in its extreme left position, while $T_{22}(v_1)$ should be moved to the extreme right position. Hereby,
the argument $v_1$ of $T_{22}$ should be preserved. Then we arrive at
\be{Lam-11}
\Lambda_1=g(u,v_1)g(\bv_1,v_1).
\ee

Similarly, calculating $\tilde \Lambda_b$ we should consider only the term with $k=b$ in \eqref{Mult-T12T231}
\begin{equation}\label{tLam-1}
[T_{12}(u),\bT_{23}(\bv)]=g(u,v_b)\frac{\bT_{23}(\bv_{b})}{h(v_b,\bv_b)}\bigl(T_{13}(u)T_{22}(v_b)-
T_{13}(v_b)T_{22}(u)\bigr)+UWT.
\end{equation}
Now $T_{22}(u)$ should be kept in its extreme right position, while $T_{13}(v_b)$ should be moved to the left preserving its argument. We find
\be{tLam-11}
\tilde\Lambda_b=-g(u,v_b)g(\bv_b,v_b).
\ee
Finally, it is easy to see that  $M_{21}=0$. Indeed, in order to obtain $M_{21}$ we should take only the term with $k=1$ in \eqref{Mult-T12T231}.
In other words we arrive at \eqref{Lam-1}. Otherwise we cannot obtain $T_{22}(v_1)$ in the extreme right position. But  the operator
$T_{13}$ is already in the extreme left position, hence, it cannot depend on $v_2$. Thus, we conclude that $M_{21}=0$.

Using now the symmetry properties we find
\be{tLamLam-1}
\Lambda_k=g(u,v_k)g(\bv_k,v_k), \qquad \tilde\Lambda_k=-g(u,v_k)g(\bv_k,v_k), \qquad M_{jk}=0,
\ee
and hence,
\begin{equation}\label{Mult-T12T23-res}
[T_{12}(u),\bT_{23}(\bv)]=\sum_{k=1}^b g(u,v_k)g(\bv_k,v_k)\Bigl(T_{13}(u)\bT_{23}(\bv_{k})\,T_{22}(v_k)-
T_{13}(v_k)\,\bT_{23}(\bv_{k})\,T_{22}(u)\Bigr),
\end{equation}
what is equivalent to \eqref{Mult-T12T230}.


\section{MCR \eqref{Y-Xab} for $a$ and $b$ arbitrary \label{S-MCRab}}

In order to show that  the operator $Y_{a,b}(\bu,\bv)$ satisfies recursion \eqref{recurs-Y} we
act with  $T_{12}(u_a)$ on the operator $Y_{a-1,b}(\bu_a,\bv)$
\be{act-T12}
T_{12}(u_a)\, Y_{a-1,b}(\bu_a,\bv)=
\sum K(\bv_{\so}|\bu_{\so}) f(\bu_{\so},\bu_{\bar\so}) g(\bv_{\bar\so},\bv_{\so})\;
T_{12}(u_a)\bT_{13}(\bv_{\so})\,\bT_{23}(\bv_{\bar\so})\,T_{12}(\bu_{\bar\so})\,T_{22}(\bu_{\so}).
\ee
Here the sum is taken over partitions $\bv\Rightarrow\{\bv_{\so},\bv_{\bar\so}\}$ and $\bu_a\Rightarrow\{\bu_{\so},\bu_{\bar\so}\}$.

The goal is to move $T_{12}(u_a)$ to the right in such a way that eventually to have all the operators in the following order:
\be{order}
\bT_{13}\;\,\bT_{23}\;\;T_{12}\;T_{22}.
\ee
The analysis of the commutation relations shows that moving $T_{12}(u_a)$ to the right we might obtain several types of the operator products
\begin{align}\label{type1}
&\bT_{13}(\bv_{\st})\,\bT_{23}(\bv_{\sth})\,T_{12}(\bu_{\sth})T_{12}(u_a)\,T_{22}(\bu_{\st}),\\
&\bT_{13}(\bv_{\st})\,\bT_{23}(\bv_{\sth})\,T_{12}(\bu_{\sth})\,T_{22}(\bu_{\st})T_{22}(u_a),\label{type2}\\
&T_{13}(u_a)\bT_{13}(\bv_{\st})\,\bT_{23}(\bv_{\sth})\,T_{12}(\bu_{\sth})\,T_{22}(\bu_{\st})T_{22}(\bv_{\qo}),\label{type3}\\
&T_{13}(u_a)\bT_{13}(\bv_{\st})\,\bT_{23}(\bv_{\sth})\,T_{12}(\bu_{\sth})T_{12}(\bv_{\qo})\,T_{22}(\bu_{\st}).\label{type4}
\end{align}
The first two types \eqref{type1} and \eqref{type2} coincide with the operator products in the definition of
$Y_{a,b}(\bu,\bv)$. We call such term {\it wanted terms} (WT). The remaining terms are unwanted (UWT). We will see that the terms
of the type \eqref{type3}  cancel the sum over partitions in \eqref{recurs-Y}, while  the terms of the
type \eqref{type4} cancel them selves.

\subsection{Wanted terms}

In this section we pay attention to the wanted terms only. The wanted terms are characterized by the property that
the operators $T_{13}$ and $T_{23}$ depend only on the variables from the set $\bv$, while the operators $T_{12}$ and $T_{22}$ depend
only on the variables from the set $\bu$.

Moving $T_{12}(u_a)$ through the product
$\bT_{13}(\bv_{\so})$ we use \eqref{TijTi3-0}. Let us write it in the form
\begin{equation}\label{Mult-T12T13uuu}
T_{12}(u)\bT_{13}(\bv_{\so})=f(\bv_{\so},u)\bT_{13}(\bv_{\so})T_{12}(u)+\sum g(u,\bv_{\qo}) g(\bv_{\qt},\bv_{\qo})T_{13}(u)\bT_{13}(\bv_{\qt})T_{12}(\bv_{\qo}),
\end{equation}
where the sum is taken over partitions $\bv_{\so}\Rightarrow\{\bv_{\qo},\bv_{\qt}\}$ with $\#\bv_{\qo}=1$. Substituting this into the
r.h.s. of \eqref{act-T12} we obtain
\begin{multline}\label{WT-1}
T_{12}(u_a)\, Y_{a-1,b}(\bu_a,\bv)=
\sum K(\bv_{\so}|\bu_{\so}) f(\bu_{\so},\bu_{\bar\so}) g(\bv_{\bar\so},\bv_{\so})
 \Bigl[f(\bv_{\so},u_a)\bT_{13}(\bv_{\so})T_{12}(u_a)\\
 + g(u_a,\bv_{\qo}) g(\bv_{\qt},\bv_{\qo})T_{13}(u_a)\bT_{13}(\bv_{\qt})T_{12}(\bv_{\qo})\Bigr]
 \bT_{23}(\bv_{\bar\so})\,T_{12}(\bu_{\bar\so})\,T_{22}(\bu_{\so}).
\end{multline}
The second term in the square brackets is unwanted, as it contains $T_{13}(u_a)$. Therefore we postpone its consideration and arrive at
\begin{multline}\label{WT-2}
T_{12}(u_a)\, Y_{a-1,b}(\bu_a,\bv)=
\sum K(\bv_{\so}|\bu_{\so}) f(\bu_{\so},\bu_{\bar\so}) g(\bv_{\bar\so},\bv_{\so})f(\bv_{\so},u_a)\\
\times\bT_{13}(\bv_{\so})T_{12}(u_a) \bT_{23}(\bv_{\bar\so})\,T_{12}(\bu_{\bar\so})\,T_{22}(\bu_{\so})+UWT.
\end{multline}

Now we use \eqref{Mult-T12T230}:
\begin{multline}\label{WT-3}
T_{12}(u_a)\, Y_{a-1,b}(\bu_a,\bv)=
\sum K(\bv_{\so}|\bu_{\so}) f(\bu_{\so},\bu_{\bar\so}) g(\bv_{\bar\so},\bv_{\so})f(\bv_{\so},u_a)\bT_{13}(\bv_{\so})\\
\times\Bigl\{g(u_a,\bv_{\qo}) g(\bv_{\qt},\bv_{\qo})\Bigl(T_{13}(u_a)\bT_{23}(\bv_{\qt})\,T_{22}(\bv_{\qo})-
T_{13}(\bv_{\qo})\,\bT_{23}(\bv_{\qt})\,T_{22}(u_a)\Bigr)\\
+\bT_{23}(\bv_{\bar\so})T_{12}(u_a)\Bigr\}
 T_{12}(\bu_{\bar\so})\,T_{22}(\bu_{\so})+UWT.
\end{multline}
Here we have an additional partition $\bv_{\bar\so}\Rightarrow\{\bv_{\qo},\bv_{\qt}\}$ with $\#\bv_{\qo}=1$.

The contribution proportional to $T_{13}(u_a)\bT_{23}(\bv_{\qt})\,T_{22}(\bv_{\qo})$ is again unwanted. The  term proportional to
$\bT_{23}(\bv_{\bar\so})T_{12}(u_a)$ already contains all the operators in the necessary order. Thus, we obtain the first
contribution to the wanted terms
\be{A1}
W_1=\sum K(\bv_{\so}|\bu_{\so}) f(\bu_{\so},\bu_{\bar\so}) g(\bv_{\bar\so},\bv_{\so})f(\bv_{\so},u_a)\bT_{13}(\bv_{\so})
\bT_{23}(\bv_{\bar\so})T_{12}(u_a)T_{12}(\bu_{\bar\so})\,T_{22}(\bu_{\so}).
\ee

It remains to consider the last contribution in \eqref{WT-3}, which is proportional to the product $T_{13}(\bv_{\qo})\,\bT_{23}(\bv_{\qt})\,T_{22}(u_a)$.
Substituting
$\bv_{\bar\so} =\{\bv_{\qt},\bv_{\qo}\}$ in this term we obtain
\begin{multline}\label{WT-4}
T_{12}(u_a)\, Y_{a-1,b}(\bu_a,\bv)-W_1=
\sum K(\bv_{\so}|\bu_{\so}) f(\bu_{\so},\bu_{\bar\so}) g(\bv_{\qt},\bv_{\so})g(\bv_{\qo},\bv_{\so})f(\bv_{\so},u_a) \\
\times g(\bv_{\qt},\bv_{\qo})g(\bv_{\qo},u_a)
\bT_{13}(\bv_{\so})T_{13}(\bv_{\qo})\,\bT_{23}(\bv_{\qt})\,T_{22}(u_a)
 T_{12}(\bu_{\bar\so})\,T_{22}(\bu_{\so})+UWT.
\end{multline}
In order to complete the calculation we should move $T_{22}(u_a)$ through the product $T_{12}(\bu_{\st})$. However, before doing this
we can take the sum over partitions $\bv_{\bar\qt}\Rightarrow\{\bv_{\so},\bv_{\qo}\}$. Indeed, we have
\be{Comb-T13}
\bT_{13}(\bv_{\so})T_{13}(\bv_{\qo})=h(\bv_{\qo},\bv_{\so})\bT_{13}(\{\bv_{\so},\bv_{\qo}\})=h(\bv_{\qo},\bv_{\so})\bT_{13}(\bv_{\bar\qt}).
\ee
Hence,
\begin{multline}\label{WT-5}
T_{12}(u_a)\, Y_{a-1,b}(\bu_a,\bv)-W_1=
\sum \Bigl[  K(\bv_{\so}|\bu_{\so}) f(\bv_{\qo},\bv_{\so})f(\bv_{\so},u_a) g(\bv_{\qo},u_a)\Bigr]f(\bu_{\so},\bu_{\bar\so}) g(\bv_{\qt},\bv_{\bar\qt})\\
\times
\bT_{13}(\bv_{\bar\qt})\,\bT_{23}(\bv_{\qt})\,T_{22}(u_a)
 T_{12}(\bu_{\bar\so})\,T_{22}(\bu_{\so})+UWT.
\end{multline}
Thus, the sum over partitions $\bv_{\bar\qt}\Rightarrow\{\bv_{\so},\bv_{\qo}\}$ involves only the rational functions in the square brackets of
\eqref{WT-5}. We have
\begin{multline}\label{Sum-RF1}
\sum_{\bv_{\bar\qt}\Rightarrow\{\bv_{\so},\bv_{\qo}\}}K(\bv_{\so}|\bu_{\so}) f(\bv_{\qo},\bv_{\so})f(\bv_{\so},u_a) g(\bv_{\qo},u_a)=
f(\bv_{\bar\qt},u_a)\sum_{\bv_{\bar\qt}\Rightarrow\{\bv_{\so},\bv_{\qo}\}}K(\bv_{\so}|\bu_{\so}) \frac{f(\bv_{\qo},\bv_{\so})}{ h(\bv_{\qo},u_a)}\\
=-f(\bv_{\bar\qt},u_a)\sum_{\bv_{\bar\qt}\Rightarrow\{\bv_{\so},\bv_{\qo}\}}K(\bv_{\so}|\bu_{\so})K(u_a-c|\bv_{\qo}) f(\bv_{\qo},\bv_{\so}),
\end{multline}
where we used $K(u_a-c|\bv_{\qo})=-1/ h(\bv_{\qo},u_a)$. We see that the
sum over partitions is reduced to  lemma~\ref{main-ident}. Hence, we arrive at
\begin{equation}\label{Sum-RF1-res}
\sum_{\bv_{\bar\qt}\Rightarrow\{\bv_{\so},\bv_{\qo}\}}K(\bv_{\so}|\bu_{\so}) f(\bv_{\qo},\bv_{\so})f(\bv_{\so},u_a) g(\bv_{\qo},u_a)=
K(\bv_{\bar\qt}|\{\bu_{\so},u_a\}),
\end{equation}
what gives us
\begin{multline}\label{WT-6}
T_{12}(u_a)\, Y_{a-1,b}(\bu_a,\bv)-W_1=
\sum  K(\bv_{\so}|\{\bu_{\so},u_a\}) f(\bu_{\so},\bu_{\bar\so}) g(\bv_{\bar\so},\bv_{\so})\\
\times
\bT_{13}(\bv_{\so})\,\bT_{23}(\bv_{\bar\so})\,T_{22}(u_a)
 T_{12}(\bu_{\bar\so})\,T_{22}(\bu_{\so})+UWT.
\end{multline}
Here we have relabeled $\bv_{\qt}$ by $\bv_{\bar\so}$ and $\bv_{\bar\qt}$ by $\bv_{\so}$.

It remains to move the operator $T_{22}(u_a)$ through the product $T_{12}(\bu_{\bar\so})$.
It can be done via \eqref{Mult-comTijTik}.
In our case this equation takes the form
\begin{equation}\label{Mult-T22T12uuu}
T_{22}(u_a)T_{12}(\bu_{\bar\so})=f(u_a,\bu_{\bar\so})T_{12}(\bu_{\bar\so})T_{22}(u_a)+
\sum g(\bu_{\qo},u_a) f(\bu_{\qo},\bu_{\qt})T_{12}(u_a)T_{12}(\bu_{\qt})T_{22}(\bu_{\qo}),
\end{equation}
where the sum is taken over partitions $\bu_{\bar\so}\Rightarrow\{\bu_{\qo},\bu_{\qt}\}$ with $\#\bu_{\qo}=1$.
Substituting this equation to \eqref{WT-6} we find
\begin{equation}\label{WT-A2A3}
T_{12}(u_a)\, Y_{a-1,b}(\bu_a,\bv)-W_1=W_2+W_3+UWT,
\end{equation}
where
\begin{equation}\label{A2}
W_2=
\sum  K(\bv_{\so}|\{\bu_{\so},u_a\}) f(\bu_{\so},\bu_{\bar\so})f(u_a,\bu_{\bar\so}) g(\bv_{\bar\so},\bv_{\so})
\bT_{13}(\bv_{\so})\,\bT_{23}(\bv_{\bar\so})\,
 T_{12}(\bu_{\bar\so})\,T_{22}(u_a)T_{22}(\bu_{\so}),
\end{equation}
and
\begin{multline}\label{A3}
W_3= \sum  K(\bv_{\so}|\{\bu_{\so},u_a\}) f(\bu_{\so},\bu_{\bar\so}) g(\bv_{\bar\so},\bv_{\so})
\, g(\bu_{\qo},u_a) f(\bu_{\qo},\bu_{\qt})\\
\times
\bT_{13}(\bv_{\so})\,\bT_{23}(\bv_{\bar\so})\,
 T_{12}(u_a)T_{12}(\bu_{\qt})T_{22}(\bu_{\qo})  T_{22}(\bu_{\so}).
\end{multline}

Observe that the contribution $W_2$ can be written in the form
\begin{equation}\label{A2-new}
W_2=\sum_{u_a\in\bu_{\so}}  K(\bv_{\so}|\bu_{\so}) f(\bu_{\so},\bu_{\bar\so}) g(\bv_{\bar\so},\bv_{\so})
\bT_{13}(\bv_{\so})\,\bT_{23}(\bv_{\bar\so})\,
 T_{12}(\bu_{\bar\so})\,T_{22}(\bu_{\so}).
\end{equation}
Here, in distinction of the original formula \eqref{act-T12} we have the sum over partitions of the complete set $\bu$ (i.e. including $u_a$). However,
we have the restriction $u_a\in\bu_{\so}$.

In \eqref{A3} we can take the sum over partitions $\bu_{\bar\qt}\Rightarrow\{\bu_{\so},\bu_{\qo}\}$. Setting there $\{\bu_{\so},\bu_{\qo}\}=\bu_{\bar\qt}$ we obtain
\begin{multline}\label{A3-1}
W_3= \sum \Bigl[ K(\bv_{\so}|\{\bu_{\so},u_a\})g(\bu_{\qo},u_a)f(\bu_{\so},\bu_{\qo})\Bigr] f(\bu_{\bar\qt},\bu_{\qt}) g(\bv_{\bar\so},\bv_{\so}) \\
\times
\bT_{13}(\bv_{\so})\,\bT_{23}(\bv_{\bar\so})\,
 T_{12}(u_a)T_{12}(\bu_{\qt}) T_{22}(\bu_{\bar\qt}),
\end{multline}

The sum of the terms in the square brackets can be computed via \eqref{Ident-1}
\be{Sum-RF2}
 \sum_{\bu_{\bar\qt}\Rightarrow\{\bu_{\so},\bu_{\qo}\}}K(\bv_{\so}|\{\bu_{\so},u_a\})g(\bu_{\qo},u_a)f(\bu_{\so},\bu_{\qo})=
 \Bigl(f(\bu_{\bar\qt},u_a)-f(\bv_{\so},u_a)\Bigr)K(\bv_{\so}|\bu_{\bar\qt}).
 \ee
Thus, we arrive at
\begin{multline}\label{A3-2}
W_3= \sum K(\bv_{\so}|\bu_{\so})\Bigl(f(\bu_{\so},u_a)-f(\bv_{\so},u_a)\Bigr) f(\bu_{\so},\bu_{\bar\so}) g(\bv_{\bar\so},\bv_{\so}) \\
\times
\bT_{13}(\bv_{\so})\,\bT_{23}(\bv_{\bar\so})\,
 T_{12}(u_a)T_{12}(\bu_{\bar\so}) T_{22}(\bu_{\so}),
\end{multline}
where we relabeled $\bu_{\bar\qt}$ by $\bu_{\so}$ and $\bu_{\qt}$ by $\bu_{\bar\so}$.

We see that
\begin{multline}\label{A1A3}
W_1+W_3=\sum K(\bv_{\so}|\bu_{\so})f(\bu_{\so},u_a) f(\bu_{\so},\bu_{\bar\so}) g(\bv_{\bar\so},\bv_{\so})
\bT_{13}(\bv_{\so})\,\bT_{23}(\bv_{\bar\so})\,
 T_{12}(u_a)T_{12}(\bu_{\bar\so}) T_{22}(\bu_{\so})\\
=\sum_{u_a\in\bu_{\bar\so}}  K(\bv_{\so}|\bu_{\so}) f(\bu_{\so},\bu_{\bar\so}) g(\bv_{\bar\so},\bv_{\so})
\bT_{13}(\bv_{\so})\,\bT_{23}(\bv_{\bar\so})\,
 T_{12}(\bu_{\bar\so})\,T_{22}(\bu_{\so}).
\end{multline}
Thus, we have again the sum over partitions of the complete set $\bu$, however,
now  the restriction is $u_a\in\bu_{\bar\so}$. Hence all together we obtain
\begin{equation}\label{A1A2A3}
\sum_{k=1}^3W_k
=\sum K(\bv_{\so}|\bu_{\so}) f(\bu_{\so},\bu_{\bar\so}) g(\bv_{\bar\so},\bv_{\so})
\bT_{13}(\bv_{\so})\,\bT_{23}(\bv_{\bar\so})\,
 T_{12}(\bu_{\bar\so})\,T_{22}(\bu_{\so})= Y_{a,b}(\bu,\bv).
\end{equation}

Thus, we have shown that the contribution of the wanted terms to the action of the operator $T_{12}(u_a)$ onto $ Y_{a-1,b}(\bu_a,\bv)$
produces the operator $Y_{a,b}(\bu,\bv)$. It remains to prove that the unwanted terms do not contribute to the final
result.

\subsection{Cancellation of unwanted terms\label{A-CUT}}

We focus now at the unwanted terms. We denote their contributions by symbols $U_k$. Later we shell split these
contributions into two groups: the terms of the type \eqref{type3} will be denoted by $Z_k$;
the terms of the type \eqref{type4} will be denoted by $\bar Z_k$.

\subsubsection{First contribution}

We start with  term proportional to $T_{13}(u_a)\bT_{23}(\bv_{\qt})\,T_{22}(\bv_{\qo})$ in \eqref{WT-3}.  Substituting there
$\bv_{\bar\so}=\{\bv_{\qt},\bv_{\qo}\}$ we obtain
\begin{multline}\label{B1-1}
U_1=
\sum K(\bv_{\so}|\bu_{\so}) f(\bu_{\so},\bu_{\bar\so}) g(\bv_{\qo},\bv_{\so})g(\bv_{\qt},\bv_{\so})f(\bv_{\so},u_a)g(u_a,\bv_{\qo}) g(\bv_{\qt},\bv_{\qo})\\
\times  \bT_{13}(\bv_{\so})T_{13}(u_a)\bT_{23}(\bv_{\qt})\,T_{22}(\bv_{\qo}) T_{12}(\bu_{\bar\so})\,T_{22}(\bu_{\so}).
\end{multline}
Moving $T_{13}(u_a)$ to the left via
\be{Mov-T13}
\bT_{13}(\bv_{\so})T_{13}(u_a)=T_{13}(u_a)\bT_{13}(\bv_{\so})\frac{h(u_a,\bv_{\so})}{h(\bv_{\so},u_a)},
\ee
we find
\begin{multline}\label{B1-2}
U_1=T_{13}(u_a)
\sum K(\bv_{\so}|\bu_{\so}) f(\bu_{\so},\bu_{\bar\so})
g(\bv_{\bar\qo},\bv_{\qo})g(\bv_{\qt},\bv_{\so})f(u_a,\bv_{\so})g(u_a,\bv_{\qo}) \\
\times  \bT_{13}(\bv_{\so})\bT_{23}(\bv_{\qt})\,T_{22}(\bv_{\qo}) T_{12}(\bu_{\bar\so})\,T_{22}(\bu_{\so}).
\end{multline}

Now we move the operator $T_{22}(\bv_{\qo})$ through the product $T_{12}(\bu_{\bar\so})$ via \eqref{Mult-T22T12}
\begin{multline}\label{B1-3}
U_1=T_{13}(u_a)
\sum K(\bv_{\so}|\bu_{\so}) f(\bu_{\so},\bu_{\bar\so}) g(\bv_{\bar\qo},\bv_{\qo})g(\bv_{\qt},\bv_{\so})f(u_a,\bv_{\so})g(u_a,\bv_{\qo}) \bT_{13}(\bv_{\so})\bT_{23}(\bv_{\qt}) \\
\times \Bigl[f(\bv_{\qo},\bu_{\bar\so}) T_{12}(\bu_{\bar\so})T_{22}(\bv_{\qo})+  g(\bu_{\qth},\bv_{\qo})f(\bu_{\qth},\bu_{\sz})
T_{12}(\bv_{\qo})T_{12}(\bu_{\sz})T_{22}(\bu_{\qth})\Bigr]\,T_{22}(\bu_{\so}).
\end{multline}
Here we have an additional sum over partitions $\bu_{\bar\so}\Rightarrow\{\bu_{\sz},\bu_{\qth}\}$ with $\#\bu_{\qth}=1$. The first term in the square brackets contributes to the
type \eqref{type3}:
\begin{multline}\label{Z1}
Z_1=T_{13}(u_a)
\sum K(\bv_{\so}|\bu_{\so}) f(\bu_{\so},\bu_{\bar\so}) g(\bv_{\bar\qo},\bv_{\qo})
g(\bv_{\qt},\bv_{\so})f(u_a,\bv_{\so})f(\bv_{\qo},\bu_{\bar\so})g(u_a,\bv_{\qo}) \\
\times \bT_{13}(\bv_{\so})\bT_{23}(\bv_{\qt})   T_{12}(\bu_{\bar\so})\,T_{22}(\bu_{\so})T_{22}(\bv_{\qo}).
\end{multline}

The second term  in the square brackets contributes to the
type \eqref{type4}. Substituting there
$\bu_{\bar\so}=\{\bu_{\sz},\bu_{\qth}\}$ we obtain
\begin{multline}\label{W1-1}
\bar Z_1=T_{13}(u_a)
\sum \Bigl[ K(\bv_{\so}|\bu_{\so}) f(\bu_{\so},\bu_{\qth}) g(\bu_{\qth},\bv_{\qo})\Bigr]       f(\bu_{\so},\bu_{\sz}) f(\bu_{\qth},\bu_{\sz})     g(\bv_{\bar\qo},\bv_{\qo})g(\bv_{\qt},\bv_{\so})
\\
\times f(u_a,\bv_{\so}) g(u_a,\bv_{\qo})
 \bT_{13}(\bv_{\so})\bT_{23}(\bv_{\qt})T_{12}(\bv_{\qo})T_{12}(\bu_{\sz})T_{22}(\bu_{\qth})\,T_{22}(\bu_{\so}).
\end{multline}
Here the sum over partitions $\bu_{\bsz}\Rightarrow\{\bu_{\so},\bu_{\qth}\}$ can be taken. The product $ f(\bu_{\so},\bu_{\sz}) f(\bu_{\qth},\bu_{\sz})$
combines into $f(\bu_{\bsz},\bu_{\sz})$,
while the sum of the terms in the square brackets is a particular case of lemma~\ref{main-ident}:
\be{W1sub-sum}
\sum_{\bu_{\bsz}\Rightarrow\{\bu_{\so},\bu_{\qth}\}} K(\bv_{\so}|\bu_{\so})g(\bu_{\qth},\bv_{\qo}) f(\bu_{\so},\bu_{\qth})=-f(\bu_{\bsz},\bv_{\qo})
K(\{\bv_{\qo}-c,\bv_{\so}\}|\bu_{\bsz}).
\ee
Substituting this into \eqref{W1-1} and relabeling $\bu_{\bsz}$ by  $\bu_{\so}$ and $\bu_{\sz}$ by  $\bu_{\bar\so}$ we find
\begin{multline}\label{W1-2}
\bar Z_1=-T_{13}(u_a)
\sum f(\bu_{\so},\bv_{\qo})
K(\{\bv_{\qo}-c,\bv_{\so}\}|\bu_{\so})
f(\bu_{\so},\bu_{\bar\so})     g(\bv_{\bar\qo},\bv_{\qo})g(\bv_{\qt},\bv_{\so})
\\
\times f(u_a,\bv_{\so}) g(u_a,\bv_{\qo})
 \bT_{13}(\bv_{\so})\bT_{23}(\bv_{\qt})T_{12}(\bv_{\qo})T_{12}(\bu_{\bar\so})T_{22}(\bu_{\so}).
\end{multline}

\subsubsection{Second contribution}

We now turn back to the term with $T_{13}(u_a)$ in \eqref{WT-1}. Substituting there
$\bv_{\so}=\{\bv_{\qt},\bv_{\qo}\}$ we obtain
\begin{multline}\label{UWT-1}
U_2=T_{13}(u_a)\sum K(\{\bv_{\qt},\bv_{\qo}\}|\bu_{\so}) f(\bu_{\so},\bu_{\bar\so})
g(\bv_{\bar\so},\bv_{\qt})      g(\bv_{\bar\qo},\bv_{\qo})g(u_a,\bv_{\qo})\\
\times  \bT_{13}(\bv_{\qt})T_{12}(\bv_{\qo}) \bT_{23}(\bv_{\bar\so})\,T_{12}(\bu_{\bar\so})\,T_{22}(\bu_{\so}).
\end{multline}
Now we move $T_{12}(\bv_{\qo})$ through the product $\bT_{23}(\bv_{\bar\so})$ via \eqref{Mult-T12T230}
\begin{multline}\label{UWT-2}
U_2=T_{13}(u_a)\sum  K(\{\bv_{\qt},\bv_{\qo}\}|\bu_{\so}) f(\bu_{\so},\bu_{\bar\so})
g(\bv_{\bar\so},\bv_{\qt})g(\bv_{\bar\qo},\bv_{\qo})g(u_a,\bv_{\qo})\\
\times  \bT_{13}(\bv_{\qt})\Bigl\{
g(\bv_{\qo},\bv_{\qth}) g(\bv_{\sz},\bv_{\qth})\Bigl[ T_{13}(\bv_{\qo})\bT_{23}(\bv_{\sz})\,T_{22}(\bv_{\qth})-
 T_{13}(\bv_{\qth})\,\bT_{23}(\bv_{\sz})\,T_{22}(\bv_{\qo})
\Bigr]\\
+ \bT_{23}(\bv_{\bar\so})\,T_{12}(\bv_{\qo})\Bigr\}\,T_{12}(\bu_{\bar\so})\,T_{22}(\bu_{\so}).
\end{multline}
Here we have an additional sum over partitions $\bv_{\bar\so}\Rightarrow\{\bv_{\sz},\bv_{\qth}\}$, where $\bv_{\qth}$ consists of one element.

The  term with the product $\bT_{23}(\bv_{\bar\so})\,T_{12}(\bv_{\qo})$ gives direct contribution to the type \eqref{type4}
\begin{multline}\label{W-2}
\bar Z_2=T_{13}(u_a)\sum  K(\{\bv_{\qt},\bv_{\qo}\}|\bu_{\so}) f(\bu_{\so},\bu_{\bar\so})
g(\bv_{\bar\so},\bv_{\qt})g(\bv_{\bar\qo},\bv_{\qo})
g(u_a,\bv_{\qo})\\
\times  \bT_{13}(\bv_{\qt})\bT_{23}(\bv_{\bar\so})\,T_{12}(\bv_{\qo})\,T_{12}(\bu_{\bar\so})\,T_{22}(\bu_{\so}).
\end{multline}
In the remaining terms of \eqref{UWT-2} we can partly take the sums over partitions.  Consider the first term in the square brackets in \eqref{UWT-2}
setting there $\bv_{\bar\so}=\{\bv_{\sz},\bv_{\qth}\}$
\begin{multline}\label{UWT-3}
U_2^{(1)}=T_{13}(u_a)\sum K(\{\bv_{\qt},\bv_{\qo}\}|\bu_{\so}) f(\bu_{\so},\bu_{\bar\so}) g(\bv_{\sz},\bv_{\qt})g(\bv_{\qth},\bv_{\qt})g(\bv_{\bar\qo},\bv_{\qo})g(u_a,\bv_{\qo})\\
\times
g(\bv_{\qo},\bv_{\qth}) g(\bv_{\sz},\bv_{\qth}) \bT_{13}(\bv_{\qt})T_{13}(\bv_{\qo})\bT_{23}(\bv_{\sz})\,T_{22}(\bv_{\qth})
\,T_{12}(\bu_{\bar\so})\,T_{22}(\bu_{\so}).
\end{multline}
In this formula the set $\bv$ is divided into four subsets $\{\bv_{\qth},\bv_{\qo},\bv_{\qt},\bv_{\sz}\}$ with $\#\bv_{\qth}=\#\bv_{\qo}=1$. Combining $\{\bv_{\qt},\bv_{\qo}\}$ into $\bv_{\sth}$ we obtain
\begin{multline}\label{UWT-4}
U_2^{(1)}=T_{13}(u_a)\sum K(\bv_{\sth}|\bu_{\so}) f(\bu_{\so},\bu_{\bar\so}) g(\bv_{\sz},\bv_{\sth})g(\bv_{\bar\qth},\bv_{\qth})
\Bigl[ f(\bv_{\qo},\bv_{\qt})g(\bv_{\qth},\bv_{\qo})g(u_a,\bv_{\qo})\Bigr]   \\
\times
  \bT_{13}(\bv_{\sth})\bT_{23}(\bv_{\sz})\,T_{22}(\bv_{\qth})
\,T_{12}(\bu_{\bar\so})\,T_{22}(\bu_{\so}).
\end{multline}
The sum over partitions $\bv_{\sth}\Rightarrow\{\bv_{\qt},\bv_{\qo}\}$  (see the terms in the square brackets) can be easily computed via contour integration:
\be{Con-int0}
\sum_{\bv_{\sth}\Rightarrow\{\bv_{\qt},\bv_{\qo}\}}f(\bv_{\qo},\bv_{\qt})g(\bv_{\qth},\bv_{\qo})g(u_a,\bv_{\qo})
=\frac1{2\pi i c}\oint f(z,\bv_{\sth})g(\bv_{\qth},z)g(u_a,z)\,dz,
\ee
where the integration contour surrounds the points $\bv_{\sth}$. Taking this integral by the residues outside the contour we find
\be{Con-int}
\sum_{\bv_{\sth}\Rightarrow\{\bv_{\qt},\bv_{\qo}\}}f(\bv_{\qo},\bv_{\qt})g(\bv_{\qth},\bv_{\qo})g(u_a,\bv_{\qo})
=g(u_a,\bv_{\qth})\bigl(f(\bv_{\qth},\bv_{\sth})-f(u_a,\bv_{\sth})\bigr).
\ee
Substituting this into \eqref{UWT-4} we obtain
\begin{multline}\label{UWT-5}
U_2^{(1)}=T_{13}(u_a)\sum K(\bv_{\sth}|\bu_{\so}) f(\bu_{\so},\bu_{\bar\so}) g(\bv_{\sz},\bv_{\sth})g(\bv_{\bar\qth},\bv_{\qth})
g(u_a,\bv_{\qth})\bigl(f(\bv_{\qth},\bv_{\sth})-f(u_a,\bv_{\sth})\bigr)\\
\times
  \bT_{13}(\bv_{\sth})\bT_{23}(\bv_{\sz})\,T_{22}(\bv_{\qth})
\,T_{12}(\bu_{\bar\so})\,T_{22}(\bu_{\so}).
\end{multline}

The second term in the square brackets of \eqref{UWT-2} can be treated similarly. Again setting  $\bv_{\bar\so}=\{\bv_{\sz},\bv_{\qth}\}$
we obtain
\begin{multline}\label{UWT-7}
U_2^{(2)}=T_{13}(u_a)\sum K(\{\bv_{\qt},\bv_{\qo}\}|\bu_{\so}) f(\bu_{\so},\bu_{\bar\so}) g(\bv_{\sz},\bv_{\qt})g(\bv_{\qth},\bv_{\qt})g(\bv_{\bar\qo},\bv_{\qo})g(u_a,\bv_{\qo})\\
\times  g(\bv_{\qth},\bv_{\qo}) g(\bv_{\sz},\bv_{\qth})
\bT_{13}(\bv_{\qt})T_{13}(\bv_{\qth})\,\bT_{23}(\bv_{\sz})\,T_{22}(\bv_{\qo})
T_{12}(\bu_{\bar\so})\,T_{22}(\bu_{\so}).
\end{multline}
This time we combine $\{\bv_{\qt},\bv_{\qth}\}$ into $\bv_{\sth}$ and find
\begin{multline}\label{UWT-8}
U_2^{(2)}=T_{13}(u_a)\sum \Bigl[K(\{\bv_{\qt},\bv_{\qo}\}|\bu_{\so}) f(\bv_{\qth},\bv_{\qt})g(\bv_{\qth},\bv_{\qo})\Bigr]
f(\bu_{\so},\bu_{\bar\so}) g(\bv_{\sz},\bv_{\sth})g(\bv_{\bar\qo},\bv_{\qo})g(u_a,\bv_{\qo})
 \\
\times
\bT_{13}(\bv_{\sth})\,\bT_{23}(\bv_{\sz})\,T_{22}(\bv_{\qo})
T_{12}(\bu_{\bar\so})\,T_{22}(\bu_{\so}).
\end{multline}
The sum over partitions  $\bv_{\sth}\Rightarrow\{\bv_{\qt},\bv_{\qth}\}$ (see the terms in the square brackets) can be computed via \eqref{Ident-2}
\be{sum-forget}
\sum_{\bv_{\sth}\Rightarrow\{\bv_{\qt},\bv_{\qth}\}}K(\{\bv_{\qt},\bv_{\qo}\}|\bu_{\so}) f(\bv_{\qth},\bv_{\qt})g(\bv_{\qth},\bv_{\qo})=
K(\bv_{\sth}|\bu_{\so})\bigl(f(\bv_{\qo},\bu_{\so})-f(\bv_{\qo},\bv_{\sth})\bigr).
\ee
Substituting this into \eqref{UWT-8}  we arrive at
\begin{multline}\label{UWT-9}
U_2^{(2)}=T_{13}(u_a)\sum K(\bv_{\sth}|\bu_{\so})\bigl(f(\bv_{\qo},\bu_{\so})-f(\bv_{\qo},\bv_{\sth})\bigr)
f(\bu_{\so},\bu_{\bar\so}) g(\bv_{\sz},\bv_{\sth})g(\bv_{\bar\qo},\bv_{\qo})g(u_a,\bv_{\qo})
 \\
\times
\bT_{13}(\bv_{\sth})\,\bT_{23}(\bv_{\sz})\,T_{22}(\bv_{\qo})
T_{12}(\bu_{\bar\so})\,T_{22}(\bu_{\so}).
\end{multline}
It remains to combine the contributions \eqref{UWT-9} and \eqref{UWT-5} (relabeling $\bv_{\qth}$ by $\bv_{\qo}$ in  \eqref{UWT-5}), what gives us
\begin{multline}\label{B2-fin}
U_2^{(1)}+U_2^{(2)}=T_{13}(u_a)\sum K(\bv_{\sth}|\bu_{\so}) f(\bu_{\so},\bu_{\bar\so}) g(\bv_{\sz},\bv_{\sth})g(\bv_{\bar\qo},\bv_{\qo})g(u_a,\bv_{\qo})
 \\
\times \bigl(f(\bv_{\qo},\bu_{\so})-f(u_a,\bv_{\sth})\bigr)
\bT_{13}(\bv_{\sth})\,\bT_{23}(\bv_{\sz})\,T_{22}(\bv_{\qo})
T_{12}(\bu_{\bar\so})\,T_{22}(\bu_{\so}).
\end{multline}

\subsubsection{Final cancellations}

To achieve our goal we should move $T_{22}(\bv_{\qo})$ in \eqref{B2-fin} through the product $T_{12}(\bu_{\bar\so})$ via \eqref{Mult-T22T12}
\begin{multline}\label{FS-1}
U_2^{(1)}+U_2^{(2)}=T_{13}(u_a)\sum K(\bv_{\sth}|\bu_{\so}) f(\bu_{\so},\bu_{\bar\so}) g(\bv_{\sz},\bv_{\sth})g(\bv_{\bar\qo},\bv_{\qo})g(u_a,\bv_{\qo})
 \\
\times \bigl(f(\bv_{\qo},\bu_{\so})-f(u_a,\bv_{\sth})\bigr)
\bT_{13}(\bv_{\sth})\,\bT_{23}(\bv_{\sz})
\Bigl[ f(\bv_{\qo},\bu_{\bar\so})T_{12}(\bu_{\bar\so})T_{22}(\bv_{\qo})\\
+g(\bu_{\qo},\bv_{\qo}) f(\bu_{\qo},\bu_{\qt})T_{12}(\bv_{\qo})T_{12}(\bu_{\qt})T_{22}(\bu_{\qo})\Bigr]
\,T_{22}(\bu_{\so}).
\end{multline}
Here we obtain additional partitions $\bu_{\bar\so}\Rightarrow\{\bu_{\qt},\bu_{\qo}\}$, where $\bu_{\qo}$ consists of one element. The first term
in the square brackets contributes to the terms of the type \eqref{type3}
\begin{multline}\label{Z2}
Z_2=T_{13}(u_a)\sum K(\bv_{\sth}|\bu_{\so}) f(\bu_{\so},\bu_{\bar\so}) f(\bv_{\qo},\bu_{\bar\so})
g(\bv_{\sz},\bv_{\sth})g(\bv_{\bar\qo},\bv_{\qo})g(u_a,\bv_{\qo})
 \\
\times \bigl(f(\bv_{\qo},\bu_{\so})-f(u_a,\bv_{\sth})\bigr)
\bT_{13}(\bv_{\sth})\,\bT_{23}(\bv_{\sz})
T_{12}(\bu_{\bar\so})\,T_{22}(\bu_{\so})T_{22}(\bv_{\qo}).
\end{multline}
Replacing here $\bv_{\sth}\to\bv_{\so}$ and $\bv_{\sz}\to\bv_{\qt}$ we obtain
\begin{multline}\label{Z2-d}
Z_2=T_{13}(u_a)\sum K(\bv_{\so}|\bu_{\so}) f(\bu_{\so},\bu_{\bar\so}) f(\bv_{\qo},\bu_{\bar\so})
g(\bv_{\qt},\bv_{\so})g(\bv_{\bar\qo},\bv_{\qo})g(u_a,\bv_{\qo})
 \\
\times \bigl(f(\bv_{\qo},\bu_{\so})-f(u_a,\bv_{\so})\bigr)
\bT_{13}(\bv_{\so})\,\bT_{23}(\bv_{\qt})
T_{12}(\bu_{\bar\so})\,T_{22}(\bu_{\so})T_{22}(\bv_{\qo}).
\end{multline}
Combining this contribution with \eqref{Z1} we find
\begin{multline}\label{FS-2}
Z_1+Z_2=T_{13}(u_a)\sum K(\bv_{\so}|\bu_{\so}) f(\bu_{\so},\bu_{\bar\so}) g(\bv_{\qt},\bv_{\so})f(\bv_{\qo},\bu_a)
g(\bv_{\bar\qo},\bv_{\qo})g(u_a,\bv_{\qo})
 \\
\times
\bT_{13}(\bv_{\so})\,\bT_{23}(\bv_{\qt})
T_{12}(\bu_{\bar\so})\,T_{22}(\bu_{\so})T_{22}(\bv_{\qo}),
\end{multline}
and hence,
\be{FS-22}
Z_1+Z_2=\sum g(u_a,\bv_{\qo})f(\bv_{\qo},\bu_a)g(\bv_{\bar\qo},\bv_{\qo})\;T_{13}(u_a)Y_{a-1,b-1}(\bu_a,\bv_{\bar\qo})T_{22}(\bv_{\qo}).
\ee
We see that this contribution cancels the sum over partitions in the recursion \eqref{recurs-Y}.

The second term in the square brackets of \eqref{FS-1}
contributes to the terms of the type \eqref{type4}, therefore we denote
it by $\bar Z_3$. We have
\begin{multline}\label{FS-3}
\bar Z_3=T_{13}(u_a)\sum K(\bv_{\sth}|\bu_{\so}) f(\bu_{\so},\bu_{\qt})f(\bu_{\so},\bu_{\qo}) g(\bv_{\sz},\bv_{\sth})
 g(\bu_{\qo},\bv_{\qo}) f(\bu_{\qo},\bu_{\qt})g(\bv_{\bar\qo},\bv_{\qo})g(u_a,\bv_{\qo})
 \\
\times \bigl(f(\bv_{\qo},\bu_{\so})-f(u_a,\bv_{\sth})\bigr)
\bT_{13}(\bv_{\sth})\,\bT_{23}(\bv_{\sz})
T_{12}(\bv_{\qo})T_{12}(\bu_{\qt})T_{22}(\bu_{\qo})
\,T_{22}(\bu_{\so}),
\end{multline}
where we substituted $\bu_{\bar\so}=\{\bu_{\qt},\bu_{\qo}\}$. Combining
$\{\bu_{\so},\bu_{\qo}\}$ into $\bu_{\sz}$ we obtain
\begin{multline}\label{FS-4}
\bar Z_3=T_{13}(u_a)\sum \Bigl[ K(\bv_{\sth}|\bu_{\so}) f(\bu_{\so},\bu_{\qo})g(\bu_{\qo},\bv_{\qo}) \Bigr]
f(\bu_{\sz},\bu_{\qt})g(\bv_{\sz},\bv_{\sth}) g(\bv_{\bar\qo},\bv_{\qo})g(u_a,\bv_{\qo})
 \\
\times \bigl( f(\bv_{\qo},\bu_{\so})-f(u_a,\bv_{\sth})\bigr)
\bT_{13}(\bv_{\sth})\,\bT_{23}(\bv_{\sz})
T_{12}(\bv_{\qo})T_{12}(\bu_{\qt})
\,T_{22}(\bu_{\sz}).
\end{multline}
We see that we can take the sum over partitions $\bu_{\sz}\Rightarrow\{\bu_{\so},\bu_{\qo}\}$.
This sum  involves the  terms in the square brackets and the product $f(\bv_{\qo},\bu_{\so})$. Thus, we have two contributions.
The first one comes from the sum
\be{LC-1}
\sum_{\bu_{\sz}\Rightarrow\{\bu_{\so},\bu_{\qo}\}}K(\bv_{\sth}|\bu_{\so})f(\bu_{\so},\bu_{\qo})g(\bu_{\qo},\bv_{\qo})=-f(\bu_{\sz},\bv_{\qo})
K(\{\bv_{\qo}-c,\bv_{\sth}\}|\bu_{\sz}),
\ee
where we have used lemma~\ref{main-ident}. The second contribution comes from the sum
\begin{multline}\label{LC-2}
\sum_{\bu_{\sz}\Rightarrow\{\bu_{\so},\bu_{\qo}\}}K(\bv_{\sth}|\bu_{\so})f(\bu_{\so},\bu_{\qo})g(\bu_{\qo},\bv_{\qo})f(\bv_{\qo},\bu_{\so})
=-f(\bv_{\qo},\bu_{\sz})\sum_{\bu_{\sz}\Rightarrow\{\bu_{\so},\bu_{\qo}\}}K(\bv_{\sth}|\bu_{\so})\frac{f(\bu_{\so},\bu_{\qo})}
{h(\bv_{\qo},\bu_{\qo})}\\
=f(\bv_{\qo},\bu_{\sz})\sum_{\bu_{\sz}\Rightarrow\{\bu_{\so},\bu_{\qo}\}}K(\bv_{\sth}|\bu_{\so})K(\bu_{\qo}|\bv_{\qo}+c)f(\bu_{\so},\bu_{\qo})=
-K(\{\bv_{\qo},\bv_{\sth}\}|\bu_{\sz}),
\end{multline}
where again we have used lemma~\ref{main-ident} and $1/h(\bv_{\qo},\bu_{\qo})=-K(\bu_{\qo}|\bv_{\qo}+c)$.

The sum \eqref{LC-1} produces the contribution
\begin{multline}\label{FS-40}
\bar Z_3^{(1)}=T_{13}(u_a)\sum \Bigl[K(\{\bv_{\qo}-c,\bv_{\sth}\}|\bu_{\bar\qt})f(\bu_{\bar\qt},\bv_{\qo})f(u_a,\bv_{\sth})\Bigr]
f(\bu_{\bar\qt},\bu_{\qt})g(\bv_{\sz},\bv_{\sth}) g(\bv_{\bar\qo},\bv_{\qo})g(u_a,\bv_{\qo})
 \\
\times
\bT_{13}(\bv_{\sth})\,\bT_{23}(\bv_{\sz})
T_{12}(\bv_{\qo})T_{12}(\bu_{\qt})
\,T_{22}(\bu_{\bar\qt}),
\end{multline}
where we used $\bu_{\sz}=\bu_{\bar\qt}$. After relabeling the subsets $\bv_{\sth}\to\bv_{\so}$, $\bv_{\sz}\to\bv_{\qt}$, $\bu_{\qt}\to\bu_{\bar\so}$, and
$\bu_{\bar\qt}\to\bu_{\so}$ we arrive at
\begin{multline}\label{FS-41}
\bar Z_3^{(1)}=T_{13}(u_a)\sum \Bigl[K(\{\bv_{\qo}-c,\bv_{\so}\}|\bu_{\so})f(\bu_{\so},\bv_{\qo})f(u_a,\bv_{\so})\Bigr]
f(\bu_{\so},\bu_{\bar\so})g(\bv_{\qt},\bv_{\so}) g(\bv_{\bar\qo},\bv_{\qo})g(u_a,\bv_{\qo})
 \\
\times
\bT_{13}(\bv_{\so})\,\bT_{23}(\bv_{\qt})
T_{12}(\bv_{\qo})T_{12}(\bu_{\bar\so})
\,T_{22}(\bu_{\so}).
\end{multline}
Comparing this equation with \eqref{W1-2} we see that $\bar Z_3^{(1)}=-\bar Z_1$.

Similarly,
the sum \eqref{LC-2} produces the term
\begin{multline}\label{FS-6}
\bar Z_3^{(2)}=-T_{13}(u_a)\sum K(\{\bv_{\qo},\bv_{\sth}\}|\bu_{\bar\qt}) f(\bu_{\bar\qt},\bu_{\qt}) g(\bv_{\sz},\bv_{\sth})
 g(\bv_{\bar\qo},\bv_{\qo})g(u_a,\bv_{\qo})
 \\
\times
\bT_{13}(\bv_{\sth})\,\bT_{23}(\bv_{\sz})
T_{12}(\bv_{\qo})T_{12}(\bu_{\qt})
\,T_{22}(\bu_{\bar\qt}).
\end{multline}
Here we again relabel the subsets $\bv_{\sth}\to\bv_{\qt}$, $\bv_{\sz}\to\bv_{\bar\so}$, $\bu_{\qt}\to\bu_{\bar\so}$, and
$\bu_{\bar\qt}\to\bu_{\so}$, what gives us
\begin{multline}\label{FS-66}
\bar Z_3^{(2)}=-T_{13}(u_a)\sum K(\{\bv_{\qo},\bv_{\qt}\}|\bu_{\so}) f(\bu_{\so},\bu_{\bar\so}) g(\bv_{\bar\so},\bv_{\qt})
 g(\bv_{\bar\qo},\bv_{\qo})g(u_a,\bv_{\qo})
 \\
\times
\bT_{13}(\bv_{\qt})\,\bT_{23}(\bv_{\bar\so})
T_{12}(\bv_{\qo})T_{12}(\bu_{\bar\so})
\,T_{22}(\bu_{\so}).
\end{multline}
Comparing this expression  with \eqref{W-2} we see that $\bar Z_3^{(2)}=-\bar Z_2$. Thus, the terms of the type \eqref{type4} do cancel them selves,
and we see that unwanted terms do not contribute to the final result.

Thus, we have proved, that the operator $Y_{a,b}(\bu,\bv)$ satisfies recursion \eqref{recurs-Y}.


\section*{Acknowledgements}
It is a great pleasure for me to thank my colleagues S.~Pakuliak and \'E.~Ragoucy for numerous
and fruitful discussions. This work was supported by the Russian Science Foundation under a grant 14-50-00005.

\appendix

\section{Properties of DWPF \label{A-ID}}

Let $\bu=\{u_1,\dots,u_n\}$ and $\bv=\{v_1,\dots,v_n\}$. The DWPF is  symmetric function of $\bu$ and symmetric function of $\bv$.
It behaves as $1/u_n$ (resp. $1/v_n$) as $u_n\to\infty$ (resp. $v_n\to\infty$) at other variables fixed.
It has simple poles at $u_j=v_k$. The behavior of $K$ near  these poles can be expressed in terms of DWPF with less number of arguments:
 \be{Rec-Ky}
\Bigl. K(\bu|\bv)\Bigr|_{u_n\to v_n}= g(u_n,v_n)
f(v_n,\bv_n)f(\bu_n,u_n)\, K(\bu_n|\bv_n)+ reg,
\ee
where $reg$ means the regular part at $u_n\to v_n$ and we recall that $\bu_n=\bu\setminus u_n$ and $\bv_n=\bv\setminus v_n$.

One can also easily check that the DWPF possesses the properties:
 \be{K-K}
 \begin{aligned}
&K(\bu, z-c|\bv, z)=K(\{\bu, z\}|\{\bv, z+c\})= - K(\bu|\bv),\\
&K(\bu-c|\bv)=K(\bu|\bv+c)= (-1)^n \frac{K(\bv|\bu)}{f(\bv,\bu) }\\
&K(\bu|\bv)\Bigr|_{c\to -c}=K(\bv|\bu).
\end{aligned}
\ee
%


\section{Summation formulas \label{A-SM}}

\subsection{Summation of Cauchy determinants}
\begin{lemma}\label{main-ident-C}
Let $\bw$, $\bu$, and $\bv$ be sets of complex variables, such that $\#\bu=m_1$,
$\#\bv=m_2$, and $\#\bw=m_1+m_2$. Then
\begin{equation}\label{Sym-Part-new1}
  \sum
 g(\bw_{\so},\bu)g(\bw_{\bar\so},\bv)g(\bw_{\bar\so},\bw_{\so})
 = \frac{g(\bw,\bu)g(\bw,\bv)}{g(\bu,\bv)}.
 \end{equation}
The sum is taken with respect to all partitions of the set $\bw$ into
subsets $\bw_{\so}$ and $\bw_{\bar\so}$, such that $\#\bw_{\so}=m_1$ and $\#\bw_{\bar\so}=m_2$.
\end{lemma}

{\sl Proof}. The proof is based on a well known explicit representation for the Cauchy determinant. Let $\#\bu=\#\bv=n$. Then
\be{det-Ca}
g(\bu,\bv)=\Delta_n(\bu)\Delta'_n(\bv)\det_n\bigl(g(u_j,v_k)\bigr).
\ee
Substituting \eqref{det-Ca} into \eqref{Sym-Part-new1} we obtain
\begin{multline}\label{proof-1}
  \sum g(\bw_{\so},\bu)g(\bw_{\bar\so},\bv)g(\bw_{\bar\so},\bw_{\so})\\
 = \sum \Delta_{m_1}(\bw_{\so})\Delta'_{m_1}(\bu)\det_{m_1}\bigl(g(w_{{\so}_j},u_k) \bigr)
 \Delta_{m_2}(\bw_{\bar\so})\Delta'_{m_2}(\bv)\det_{m_2}\bigl(g(w_{{\bar\so}_j},v_k)\bigr) g(\bw_{\bar\so},\bw_{\so}).
 \end{multline}
Obviously,
\be{D-DDD}
\Delta_{m_1}(\bw_{\so})\Delta_{m_2}(\bw_{\bar\so})g(\bw_{\bar\so},\bw_{\so})= (-1)^{[P_{{\so},{\bar\so}}]}\Delta_{m_1+m_2}(\bw),
\ee
where $P_{{\so},{\bar\so}}$ is the permutation mapping the union $\{\bw_{\so},\bw_{\bar\so}\}$ of the naturally ordered subsets $\bw_{\so}$
and $\bw_{\bar\so}$ into the naturally ordered set $\{w_1,\dots,w_{m_1+m_2}\}$. Then we arrive at
\begin{multline}\label{proof-2}
  \sum g(\bw_{\so},\bu)g(\bw_{\bar\so},\bv)g(\bw_{\bar\so},\bw_{\so})\\
 = \Delta_{m_1+m_2}(\bw)\Delta'_{m_1}(\bu)\Delta'_{m_2}(\bv)\sum (-1)^{[P_{{\so},{\bar\so}}]}\det_{m_1}\bigl(g(w_{{\so}_j},u_k) \bigr)
 \det_{m_2}\bigl(g(w_{{\bar\so}_j},v_k)\bigr) .
 \end{multline}
The obtained sum is nothing but a development of a determinant with respect to $m_1$ columns:
\be{proof-3}
\sum (-1)^{[P_{{\so},{\bar\so}}]}\det_{m_1}\bigl(g(w_{{\so}_j},u_k) \bigr)
 \det_{m_2}\bigl(g(w_{{\bar\so}_j},v_k)\bigr)=\det_{m_1+m_2}\bigl(g(w_j,\gamma_k)\bigr),
 \ee
where $\bg=\{\bu,\bv\}=\{u_1,\dots,u_{m_1},v_1,\dots,v_{m_2}\}$. Thus, we obtain
\begin{equation}\label{proof-4}
  \sum g(\bw_{\so},\bu)g(\bw_{\bar\so},\bv)g(\bw_{\bar\so},\bw_{\so})
 = \Delta_{m_1+m_2}(\bw)\Delta'_{m_1}(\bu)\Delta'_{m_2}(\bv) \det_{m_1+m_2}\bigl(g(w_j,\gamma_k)\bigr).
 \end{equation}
It remains to use \eqref{det-Ca} for the determinant of $g(w_j,\gamma_k)$:
\be{det-Cag}
\det_{m_1+m_2}\bigl(g(w_j,\gamma_k)\bigr)=\frac{g(\bw,\bu)g(\bw,\bv)}{\Delta_{m_1+m_2}(\bw)\Delta_{m_1+m_2}(\{\bu,\bv\})}.
\ee
Substituting this into \eqref{proof-4} we immediately arrive at \eqref{Sym-Part-new1}.

\begin{lemma}\label{main-ident}
Let $\bw$, $\bu$, and $\bv$ be sets of complex variables, such that $\#\bu=m_1$,
$\#\bv=m_2$, and $\#\bw=m_1+m_2$. Then
\begin{equation}\label{Sym-Part-old1}
  \sum
 K(\bw_{\so}|\bu)K(\bv|\bw_{\bar\so})f(\bw_{\bar\so},\bw_{\so})
 = (-1)^{m_1}f(\bw,\bu) K_{m_1+m_2}(\{\bu-c,\bv\}|\bw).
 \end{equation}
The sum is taken with respect to all partitions of the set $\bw$ into
subsets $\bw_{\so}$ and $\bw_{\bar\so}$, such that $\#\bw_{\so}=m_1$ and $\#\bw_{\bar\so}=m_2$.
\end{lemma}
The proof of this Lemma is given in \cite{BelPRS12a}.


\begin{lemma}\label{New-sum}
Let $\#\bu=\#\bv=n$. Then for arbitrary complex $\xi$
\be{Ident-1}
\sum K(\bv|\{\bu_{\bar\so},\xi\})g(\bu_{\so},\xi)f(\bu_{\bar\so},\bu_{\so})= \bigl(f(\bu,\xi)-f(\bv,\xi)\bigr)K(\bv|\bu).
\ee
Here the sum is taken over partitions $\bu\Rightarrow\{\bu_{\so},\bu_{\bar\so}\}$ with $\#\bu_{\so}=1$.
\end{lemma}

{\sl Proof.} Since both parts of \eqref{Ident-1} decrease at $\xi\to\infty$, it is enough to consider residues
in the poles at $\xi=u_j$ and $\xi=v_j$. In the r.h.s. they are
quite obvious. If $\xi=u_j$ in the l.h.s., then the pole occurs if and only if $u_j\in\bu_{\so}$. Since
$\#\bu_{\so}=1$, we conclude that $\bu_{\so}=u_j$, and there is only one term in the sum over partitions. Thus, we arrive at
\be{Proof-1}
\sum K(\bv|\{\bu_{\bar\so},\xi\})g(\bu_{\so},\xi)f(\bu_{\bar\so},\bu_{\so})\Bigr|_{\xi\to u_j}= g(u_j,\xi)f(\bu_j,u_j)K(\bv|\bu),
\ee
what coincides with the residue in the r.h.s.

If $\xi=v_j$, then $K(\bv|\{\bu_{\bar\so},\xi\})$ has a pole, and we use recursion \eqref{Rec-Ky}:
\begin{multline}\label{Proof-2}
\sum K(\bv|\{\bu_{\bar\so},\xi\})g(\bu_{\so},\xi)f(\bu_{\bar\so},\bu_{\so})\Bigr|_{\xi\to v_j}\\
= g(v_j,\xi)f(\bv_j,v_j)\sum f(v_j,\bu_{\bar\so})K(\bv_j|\bu_{\bar\so})g(\bu_{\so},v_j)f(\bu_{\bar\so},\bu_{\so})+reg.
\end{multline}
Now the sum over partitions can be calculated via  lemma~\ref{main-ident}
\begin{multline}\label{Proof-3}
\sum f(v_j,\bu_{\bar\so})K(\bv_j|\bu_{\bar\so})g(\bu_{\so},v_j)f(\bu_{\bar\so},\bu_{\so})=
-f(v_j,\bu)\sum K(\bv_j|\bu_{\bar\so})\frac{f(\bu_{\bar\so},\bu_{\so})}{h(v_j,\bu_{\so})}\\
=f(v_j,\bu)\sum K(\bv_j|\bu_{\bar\so})K(\bu_{\so}|v_j+c)  f(\bu_{\bar\so},\bu_{\so})
=- K(\bv|\bu),
\end{multline}
where we used $1/h(v_j,\bu_{\so})=-K(\bu_{\so}|v_j+c)$. Hence,
\be{Proof-4}
\sum K(\bv|\{\bu_{\bar\so},\xi\})g(\bu_{\so},\xi)f(\bu_{\bar\so},\bu_{\so})\Bigr|_{\xi\to v_j}
=-g(v_j,\xi)f(\bv_j,v_j)K(\bv|\bu)+reg,
\ee
what coincides with the residue in the r.h.s. of \eqref{Ident-1}.

\begin{cor}\label{New-sumd}
Under the conditions of lemma~\ref{New-sum}
\be{Ident-2}
\sum K(\{\bv_{\bar\so},\xi\}|\bu)g(\bv_{\so},\xi)f(\bv_{\so},\bv_{\bar\so})= \bigl(f(\xi,\bu)-f(\xi,\bv)\bigr)K(\bv|\bu).
\ee
Here the sum is taken over partitions $\bv\Rightarrow\{\bv_{\so},\bv_{\bar\so}\}$ with $\#\bv_{\so}=1$.
\end{cor}

This identity follows from \eqref{Ident-1} after the replacements
$c$ by $-c$ and   $\bu \leftrightarrow \bv$.


\begin{thebibliography}{99}
%
%
\bibitem{FadST79} L. D. Faddeev, E. K. Sklyanin and L. A. Takhtajan, \textsl{Quantum Inverse Problem. I},
 Theor. Math. Phys. {\bf 40} (1979) 688--706.
%
\bibitem{FadT79} L. D. Faddeev and L. A. Takhtajan, \textsl{The quantum method of the inverse problem and the Heisenberg $XYZ$ model},
Usp. Math. Nauk {\bf 34} (1979) 13;  Russian Math. Surveys {\bf 34} (1979) 11 (Engl. transl.).
%
\bibitem{FadLH96} L. D. Faddeev, in: Les Houches Lectures \textsl{Quantum Symmetries}, eds A. Connes
et al, North Holland, (1998) 149.
%
\bibitem{BogIK93L}V. E. Korepin, N. M. Bogoliubov,
A. G. Izergin, {\sl Quantum Inverse Scattering Method and Correlation Functions}, Cambridge: Cambridge Univ.
Press, 1993.
%
\bibitem{KitMT00} N.~Kitanine, J.~M. Maillet, V.~Terras, \textsl{Correlation functions of the $XXZ$ Hei\-sen\-berg spin-$1/2$ chain in a magnetic field},
Nucl. Phys. B {\bf 567} (2000) 554--582, \texttt{arXiv:math-ph/9907019}.
%
\bibitem{KitKMST12} N. Kitanine, K. Kozlowski, J. M. Maillet, N. A. Slavnov, V. Terras,
\textsl{Form factor approach to dynamical correlation functions in critical models},
J. Stat. Mech. \textbf{1209} (2012) P09001, \texttt{arXiv:1206.2630}.
%
%
\bibitem{GohKS04} F. G\"ohmann, A. Kl\"umper, A. Seel, \textsl{Integral representations for correlation
 functi\-ons of the $XXZ$ chain at finite temperature}, J. Phys. A {\bf 37} (2004) 7625--7652, \texttt{arXiv:hep-th/0405089}.
%
\bibitem{IzeK84}
 A. G. Izergin and V. E. Korepin, \textsl{The Quantum Inverse Scattering Method Approach to
Correlation Functions}, Commun. Math. Phys. {\bf 94} (1984) 67--92.
%
\bibitem{KitMST02} N. Kitanine, J.M. Maillet, N.A. Slavnov and V. Terras, \textsl{Spin-spin correlation functions of the  $XXZ$-$1/2$
Heisenberg chain in a magnetic field}, Nucl. Phys. B {\bf 641} (2002) 487--518, \texttt{arXiv:hep-th/0201045}.
%
\bibitem{BelPRS12c} S. Belliard, S. Pakuliak, E. Ragoucy, N. A. Slavnov,
{\sl Bethe vectors of $GL(3)$-invariant integrable models}, J. Stat. Mech. (2013) P02020, \texttt{arXiv:1210.0768}.
%
\bibitem{ZhaR88} F.C. Zhang, T.M. Rice, \textsl{Effective Hamiltonian for the superconducting $Cu$ oxides}, Phys. Rev. B {\bf 37} (1988) 3759--3761.
%
\bibitem{Sch87} P. Schlottmann, \textsl{Integrable narrow-band model with possible relevance to heavy
Fermion systems}, Phys. Rev. B {\bf 36} (1987) 5177--5185.
%
\bibitem{BarB90} P.A. Bares and G. Blatter, \textsl{Supersymmetric t-J model in one dimension: Separation of spin and charge},
Phys. Rev. Lett. {\bf 64}  (1990) 2567--2570.
%
\bibitem{Sar90} S. Sarkar, \textsl{Bethe-ansatz solution of the t-J model}, J. Phys. A {\bf 23} (1990) L409--L414.
%
\bibitem{BarBO91} P.A. Bares, G. Blatter, M. Ogata,
\textsl{Exact solution of the t-J model in one dimension at $2t=\pm J$: Ground state and excitation spectrum},
 Phys. Rev. B {\bf 44} (1991) 130--154.
%
\bibitem{Sar91} S. Sarkar, \textsl{The supersymmetric t-J model in one dimension}, J. Phys. A {\bf 24} (1991) 1137--1152.
%
%
\bibitem{EssK92} F.H.L. Essler and V. E. Korepin, \textsl{Higher conservation laws and algebraic Bethe
Ansatze for the supersymmetric t-J model}, Phys. Rev. B {\bf 46} (1992) 9147--9162.
%
\bibitem{FoeK93} A. Foerster and M. Karowski, \textsl{Algebraic properties of the Bethe ansatz for an
$spl(2,1)$-supersymmetric t-J model}, Nucl. Phys. B {\bf 396} (1993) 611--638..
%
\bibitem{KhoP05} S. Khoroshkin, S. Pakuliak, \textsl{Weight function for $U_q(\hat{\mathfrak{sl}}_3)$} Theor. Math.
Phys. {\bf 145}:1 (2005) 1373--1399, \texttt{arXiv:math/0610433 }.
%
\bibitem{KhoPT} S. Khoroshkin, S. Pakuliak, V. Tarasov, \textsl{Off-shell Bethe vectors and Drinfeld currents},
J. Geom. Phys. {\bf 57}:8 (2007) 1713--1732, \texttt{arXiv:math/0610517}
%
\bibitem{FraKPR09} L. Frappat, S. Khoroshkin, S. Pakuliak, E. Ragoucy,
\textsl{Bethe Ansatz for the Universal Weight Function},
Ann. H. Poincar\'e {\bf 10} (2009) 513--548, \texttt{arXiv:0810.3135}.
%
\bibitem{BelPR10} S. Belliard, S. Pakuliak, E. Ragoucy, \textsl{Bethe Ansatz and Bethe Vectors Scalar Products},
SIGMA {\bf 6} (2010) 094, \texttt{arXiv:1012.1455}.
%
\bibitem{TarV13}
V. Tarasov and A. Varchenko, \textsl{Combinatorial formulae for
nested Bethe vector}, SIGMA {\bf 9} (2013) 048 and \texttt{arXiv:math.QA/0702277}.
%
\bibitem{BelR08}
S.~Belliard and {\'E}.~Ragoucy,
\textsl{The nested Bethe ansatz for 'all' closed spin chains}, J. Phys. A {\bf 41} (2008) 295202,
\texttt{arXiv:math-ph/0804.2822}.
%
\bibitem{PakRS16a} S. Pakuliak, E. Ragoucy, N.A. Slavnov, \textsl{Bethe vectors for models based on the super-Yangian $Y(\mathfrak{gl}(m|n))$},
\texttt{arXiv:1604.02311}.
%
\bibitem{KulS80} P.P. Kulish and E.K. Sklyanin, \textsl{On the solution of the Yang-Baxter equation},
Zap. Nauchn. Semin. LOMI {\bf 95} (1980) 129--160;  J. Sov. Math. {\bf 19} (1982) 1596 (Engl. transl.).
%
\bibitem{Kor82} V. E. Korepin, \textsl{Calculation of norms of Bethe wave functions}, Comm. Math. Phys. {\bf 86} (1982) 391--418.
%
\bibitem{Ize87} A. G. Izergin, \textsl{Partition function of the six-vertex model in a finite volume},
Dokl. Akad. Nauk SSSR {\bf 297} (1987) 331--333;
Sov. Phys. Dokl. {\bf 32} (1987) 878--879 (Engl. transl.).
%
\bibitem{BelPRS12a} S. Belliard, S. Pakuliak, E. Ragoucy, N.A. Slavnov,
\textsl{Highest coefficient of scalar products in $SU(3)$-invariant integrable models}, J. Stat. Mech. Theory Exp. (2012) P09003, \texttt{arXiv:1206.4931}.



\end{thebibliography}
\end{document}